\documentclass[preprint,12pt]{elsarticle}

\usepackage{amssymb}
\usepackage{amsmath}
\usepackage{subcaption}

\usepackage{color}

\lccode`\3`\3
\lccode`\/`\/
\journal{Astroparticle Physics}

\begin{document}

\begin{frontmatter}



\title{Sensitivity of the KM3NeT/ARCA neutrino telescope to point-like neutrino sources}


\cortext[cor]{corresponding author}

\author[1]{S.~Aiello}
\author[2]{S.\,E.~Akrame}
\author[3]{F.~Ameli}
\author[4]{E.~G.~Anassontzis}
\author[5]{M.~Andre}
\author[6]{G.~Androulakis}
\author[7]{M.~Anghinolfi}
\author[8]{G.~Anton}
\author[9]{M.~Ardid}
\author[10]{J.~Aublin}
% ...
\address[1]{INFN, Sezione di Catania, Via Santa Sofia 64, Catania, 95123 Italy}
\address[2]{Cadi Ayyad University, Physics Department, Faculty of Science Semlalia, Av. My Abdellah, P.O.B. 2390, Marrakech, 40000 Morocco}
\address[3]{INFN, Sezione di Roma, Piazzale Aldo Moro 2, Roma, 00185 Italy}
\address[4]{Physics~Department,~N.~and~K.~University~of~Athens,~Athens,~Greece}
\address[5]{Universitat Polit{\`e}cnica de Catalunya, Laboratori d'Aplicacions Bioac{\'u}stiques, Centre Tecnol{\`o}gic de Vilanova i la Geltr{\'u}, Avda. Rambla Exposici{\'o}, s/n, Vilanova i la Geltr{\'u}, 08800 Spain}
\address[6]{NCSR Demokritos, Institute of Nuclear and Particle Physics, Ag. Paraskevi Attikis, Athens, 15310 Greece}
\address[7]{INFN, Sezione di Genova, Via Dodecaneso 33, Genova, 16146 Italy}
\address[8]{Friedrich-Alexander-Universit{\"a}t Erlangen-N{\"u}rnberg, Erlangen Centre for Astroparticle Physics, Erwin-Rommel-Stra{\ss}e 1, 91058 Erlangen, Germany}
\address[9]{Universitat Polit{\`e}cnica de Val{\`e}ncia, Instituto de Investigaci{\'o}n para la Gesti{\'o}n Integrada de las Zonas Costeras, C/ Paranimf, 1, Gandia, 46730 Spain}
\address[10]{APC, Universit{\'e} Paris Diderot, CNRS/IN2P3, CEA/IRFU, Observatoire de Paris, Sorbonne Paris Cit\'e, 75205 Paris, France}
% ...

\begin{abstract}
KM3NeT will be a network of deep-sea neutrino telescopes in the Mediterranean
Sea. The KM3NeT/ARCA detector, to be installed at the Capo Passero site (Italy),
is optimised for the detection of high-energy neutrinos of cosmic origin. Thanks
to its geographical location on the Northern hemisphere, KM3NeT/ARCA can observe
upgoing neutrinos from most of the Galactic Plane, including the Galactic
Centre. Given its effective area and excellent pointing resolution, KM3NeT/ARCA
will measure or significantly constrain the neutrino flux from potential
astrophysical neutrino sources. At the same time, it will test flux predictions based on
gamma-ray measurements and the assumption that the gamma-ray flux is of
hadronic origin. Assuming this scenario, discovery potentials and sensitivities for a selected list of Galactic sources and to generic point sources with an
$E^{-2}$ spectrum are presented. These spectra are assumed to be time independent. The results indicate that an
observation with $3\sigma$ significance is possible in about six years of operation for the most intense
sources, such as Supernovae Remnants RX\,J1713.7-3946 and Vela Jr. If no signal will be found during
this time, the fraction of the gamma-ray flux coming from hadronic processes can
be constrained to be below 50\% for these two objects.
\end{abstract}


\end{frontmatter}



\section{Introduction}\label{sec:intro}

Neutrinos are an optimal probe to observe high energy astrophysical phenomena,
since they interact only weakly with matter and are not deflected by magnetic
fields. Therefore, they point back to their origin, can bridge large distances
without absorption, and may provide information on processes in dense sources, which can be opaque to the electromagnetic radiation. 
They are unique messengers from the most violent and highest energy processes
in our Galaxy and far beyond. The discovery by the IceCube Collaboration
of a high-energy neutrino flux of extra-terrestrial origin \cite{icecube-2eve, icecube-28eve, IceCube37} has
thus opened a new observational window on our Universe and initiated a new era
of neutrino astronomy. 
KM3NeT\footnote{http://www.km3net.org} is a large research infrastructure that will consist of a
network of deep-sea neutrino telescopes in the Mediterranean Sea. KM3NeT will
include two detectors with the same technology but different granularity,
KM3NeT/ARCA and KM3NeT/ORCA (Astroparticle and Oscillation Research with Cosmics in the
Abyss, respectively) \cite{loi}. While KM3NeT/ORCA, installed at the KM3NeT-France site offshore Toulon
(France), will study oscillations of atmospheric neutrinos with the primary
objective to determine the neutrino mass ordering, KM3NeT/ARCA will be dedicated to
high-energy neutrino astronomy, including the investigation of the
cosmic neutrino flux discovered by IceCube. KM3NeT/ARCA is being installed at the
KM3NeT-Italy site offshore Capo Passero (Italy) and will have cubic-kilometer scale
size, suited to measure neutrinos in the TeV--PeV energy range. KM3NeT/ARCA will have a wider and complementary field of view with respect to IceCube. 
One of its primary targets is the detection of Galactic sources visible also at relatively low energy around tens of TeV for which the IceCube sensitivity to muon neutrinos
is low.

In KM3NeT, neutrinos are detected by measuring the Cherenkov light induced by charged
secondary particles emerging from a neutrino interaction in the sea water, which
serves as target material and Cherenkov radiator as well as a shield for downgoing atmospheric muons. 
The light is detected by
photo-multiplier tubes (PMTs) arranged in glass spheres that withstand the water
pressure (digital optical modules, DOMs \cite{BRUIJN, DomTest}). 
Each optical
module houses 31 3-inch PMTs 
optimising the photo-cathode area, the directional
sensitivity, the angular coverage per DOM, and the photon counting capability. 
The DOMs of the KM3NeT/ARCA detector are arranged along flexible strings with a total height of about
700\,m. KM3NeT/ARCA will consist of two building blocks of 115~strings
each, with 18~DOMs per string, vertically spaced by 36\,m. Each block will have a
roughly circular footprint with an average distance between strings of about 90\,m. The
two blocks together will cover an instrumented volume of about 1\,km$^3$. They will be
deployed and anchored in the Capo Passero site located at 36$^\circ$ 16' N 16$^\circ$  06' E, at a depth of 3500\,m, and will be connected to the
shore station via a 100 km electro-optical cable to transfer power and data between shore and the detector.

Different populations of Galactic astrophysical objects have been proposed as production sites of neutrinos up to the TeV--PeV range. 
Supernova Remnants (SNRs) are the best motivated candidates in
our Galaxy \cite{SNRneutrini}. They are often addressed as the main contributors to the flux of Galactic Cosmic Rays (the so-called SNR paradigm on the origin of GCR). 
Evidence for the acceleration of protons in the remnants was provided in 2013 when Fermi-LAT reported an indication of the pion-decay signature from the
SNRs W 44 and IC 443 \cite{FermiPioni}. However,
being model dependent, this measurement is not a conclusive proof. 
Being a smoking gun for hadronic acceleration, neutrinos could contribute to the challenge of unveiling cosmic-ray accelerators.

In the last decades, very high energy (VHE: $E_\gamma>100$\,GeV) emission from a large number of
 Galactic SNRs has been identified by $\gamma$-ray telescopes. The observed
$\gamma$-ray spectra can extend up to tens of TeV, proving that these objects are efficient particle accelerators.
These particles could be protons yielding
$\gamma$-rays via inelastic production of neutral pions, but could also be
electrons which emit VHE $\gamma$-rays via
Inverse Compton scattering on ambient low energy photons. The observation of
high-energy neutrinos from these sources would establish an unambiguous proof
that hadronic processes are at work; due to strong model dependences, this proof
cannot easily be achieved with the current $\gamma$-rays observations.

Another class of Galactic objects observed in TeV $\gamma$-rays comprises Pulsar Wind
Nebulae (PWNe), in which emission of non-thermal radiation is powered by the
relativistic outflows from a pulsar, i.e.\ a rapidly spinning, strongly
magnetised neutron star. 
The interaction of the pulsar wind with the slower supernova ejecta or with the
interstellar medium creates a termination shock where particles can be
accelerated to very high energies. Even though the TeV emission of PWN is
usually interpreted in a purely leptonic scenario \cite{PWNleptonic}, some authors \cite{horns} also consider
the presence of a hadronic contribution, which could be tested with neutrino telescopes.

The scientific potential of KM3NeT/ARCA to detect neutrino
point-like sources in our Galaxy and beyond is discussed in this paper. This subject has already been
covered in Ref.\ \cite{loi}. However, since then, the event reconstruction and
analysis methods have been improved significantly, leading to new results
presented in this paper. 
Moreover, the recent publication of new and more precise $\gamma$-ray observations \cite{HESS16, VelaJrHESS2016} has also allowed for updated neutrino flux predictions. 
In addition, an extended set of potential neutrino
sources is now investigated, including several candidate sources for measurable
neutrino signals. A stacking analysis of SNRs with the most intense VHE $\gamma$-ray flux is also presented.

The recent detection by IceCube of a high-energy neutrino event coincident in direction and time with a $\gamma$-ray flaring state of the blazar TXS 0506+056 is reported in Ref.\ \cite{IceCubeMultiMess}. 
This observation suggests that  blazars  \cite{blazar} are likely sources of extra-galactic high-energy neutrinos.
In addition, an investigation over the full IceCube neutrino archive has shown an excess with more than 3$\sigma$ significance of high-energy neutrino events at the position of this blazar compatible with a neutrino flux with $E^{-2}$ energy dependence \cite{IceCubeBlazar}.
ANTARES has also searched for neutrinos from this source \cite{ANTARESblazar} but no evidence has been found.
To illustrate the detection capabilities of KM3NeT/ARCA for this type of extragalactic sources,
 the sensitivity of KM3NeT/ARCA to a $E^{-2}$ neutrino flux from a point-like source is also discussed.

The analysis focusses on charged-current interactions of muon-neutrinos,
producing a high-energy muon in the final state. Due to its path length of up to
several kilometres in water, the direction of the muon and thus -- at
sufficiently high neutrino energy -- the neutrino direction can be
measured with good accuracy (see Section~\ref{sec:reco}). Such track-like events therefore provide the dominant contribution to
the sensitivity for point-like sources \cite{loi}. The main backgrounds are due to
atmospheric neutrinos and muons produced by the interaction of cosmic rays with nuclei in the atmosphere. To eliminate atmospheric muons, only events
reconstructed as upgoing or coming from slightly above the horizon are
selected since the Earth or the slant water layer traversed absorbs all particles
except neutrinos. The cosmic neutrino signal is observed as an excess on the
background of atmospheric neutrinos and of remaining atmospheric muons falsely
reconstructed as upgoing. Given the latitude of the detector, KM3NeT will
detect upgoing neutrinos from about $3.5\pi$\,sr of the sky, including most
of the Galactic Plane. The visibility of a given candidate source (i.e.\ the
fraction of time it is observable) depends on its declination, $\delta$, and on
the angular acceptance above the horizon, see Fig.~\ref{visibility}. In
particular, note that the region of full visibility extends to
$\delta\lesssim-45^\circ$ if events up to $10^\circ$ above the horizon are
included, as in the present analysis (see Section~\ref{sec:analysis}).

In this paper, neutrino fluxes expected from a selected list of Galactic $\gamma$-ray sources
are estimated assuming a hadronic scenario for the $\gamma$-ray production and
transparent sources. This topic is discussed in Section~\ref{sec:nuflux}. The
details of the simulation codes are described in Section~\ref{sec:simul} and the
reconstruction performances in Section~\ref{sec:reco}. The analysis procedure and
the results are presented in Section~\ref{sec:analysis}. Section~\ref{sec:E-2} is
devoted to the KM3NeT/ARCA sensitivity and discovery potential for a generic $E^{-2}$
flux. The effect of systematic uncertainties is discussed in
Section~\ref{sec:systematics}, and the conclusions are summarised in
Section~\ref{sec:conclusion}.

\section{Selected Galactic sources and estimated neutrino fluxes}
\label{sec:nuflux}

\begin{figure}[h]
\centering
\includegraphics[width=0.7\textwidth]{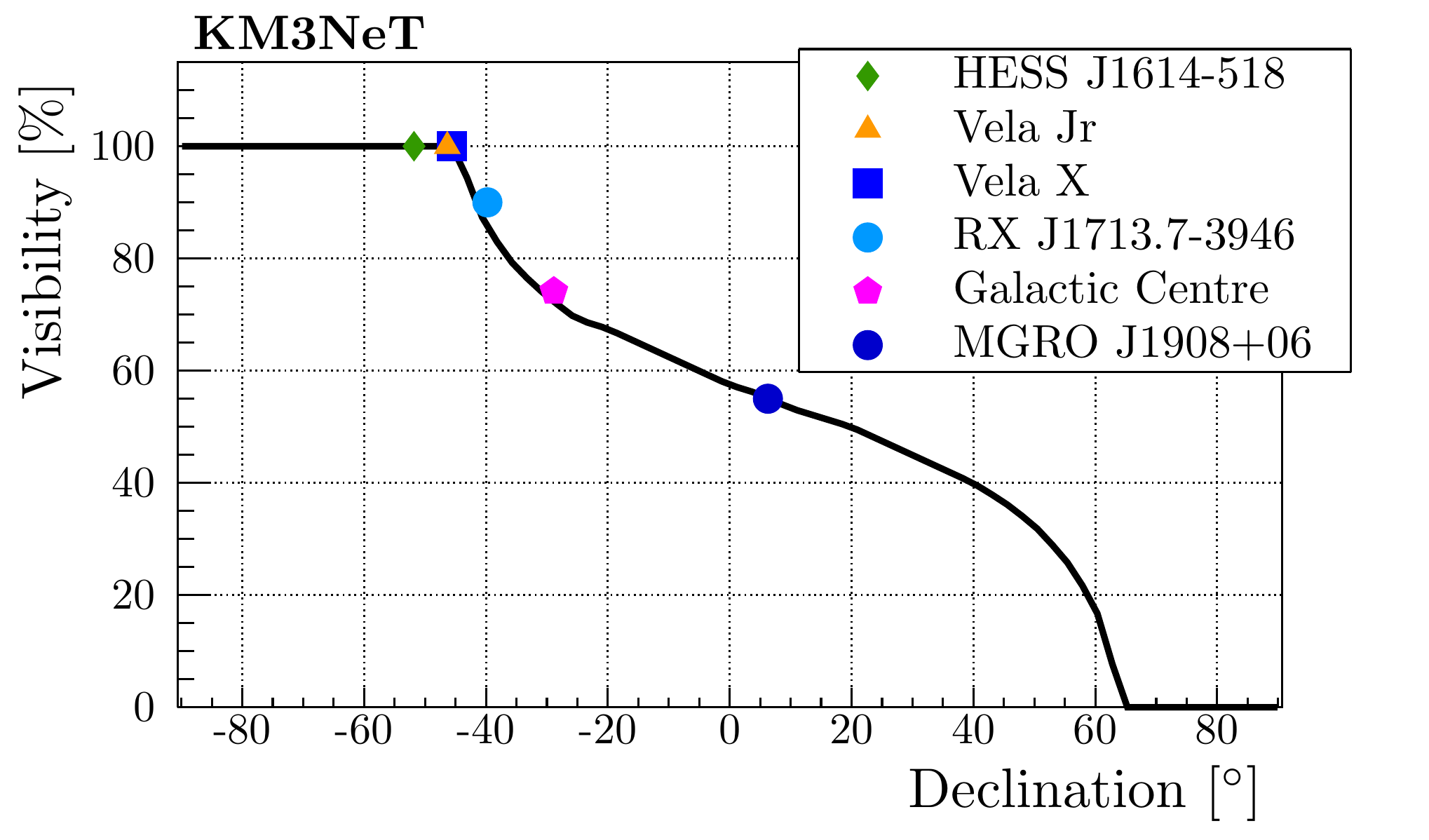}
\caption{Source visibility for KM3NeT/ARCA as a function of declination for a zenith cut of $10^\circ$ above the horizon (black line). The markers represent the visibility of the specific sources discussed in this paper according to their declination and the zenith cuts used in the analyses (see Table~\ref{tab:events} for the individual zenith cuts).}
\label{visibility}
\end{figure}

Galactic candidate sources have been selected from the TeVCat catalogue \cite{tevcat} on the basis of their
visibility, the $\gamma$-ray intensity and the energy spectrum. In particular, it was required that the $\gamma$-ray flux is measured up to a few tens of TeV.
The selected sources are: RX\,J1713.7-3946, Vela~X, Vela Jr, HESS\,J1614-518, the Galactic Centre and
MGRO\,J1908+06 (see Table~\ref{tab:source} for the individual references). 
The visibility of these sources is indicated in Fig.~\ref{visibility}. 
Except for MGRO\,J1908+06, all the sources have a visibility above 70\%.

For all the sources (with the only exception of MGRO\,J1908+06), the neutrino flux is derived from the measured $\gamma$-ray flux using the method described in Refs.\ \cite{Vissani_RXJ1713, VissaniGC} and references therein. Another method has been tested \cite{kappes}, using as a test case the source RX\,J1713.7-3946, obtaining compatible results. 
All neutrino fluxes are estimated for the $\nu_\mu$\footnote{in this paper the notation $\nu$ is used to refer to both neutrinos and antineutrinos}
channel, assuming that, due to oscillation, for cosmic neutrinos the flavour ratio at Earth will be $\nu_e:\nu_\mu:\nu_\tau=1:1:1$ \cite{Athar}. 
For all cases a 100\% hadronic emission and a transparent source are assumed, but the results can also be interpreted in terms of the percentage of hadronic emission, provided that the hadronic and non-hadronic contributions have the same energy spectrum.
If $\xi_\text{had}$ is the percentage of the $\gamma$-ray flux that has hadronic origin, the neutrino fluxes are calculated under the hypothesis that $\xi_\text{had} = 1$, but from these results the discovery potentials and sensitivities for $\xi_\text{had} < 1$ can be derived.

\begin{figure}[ht]
\centering
\includegraphics[width=0.6\textwidth]{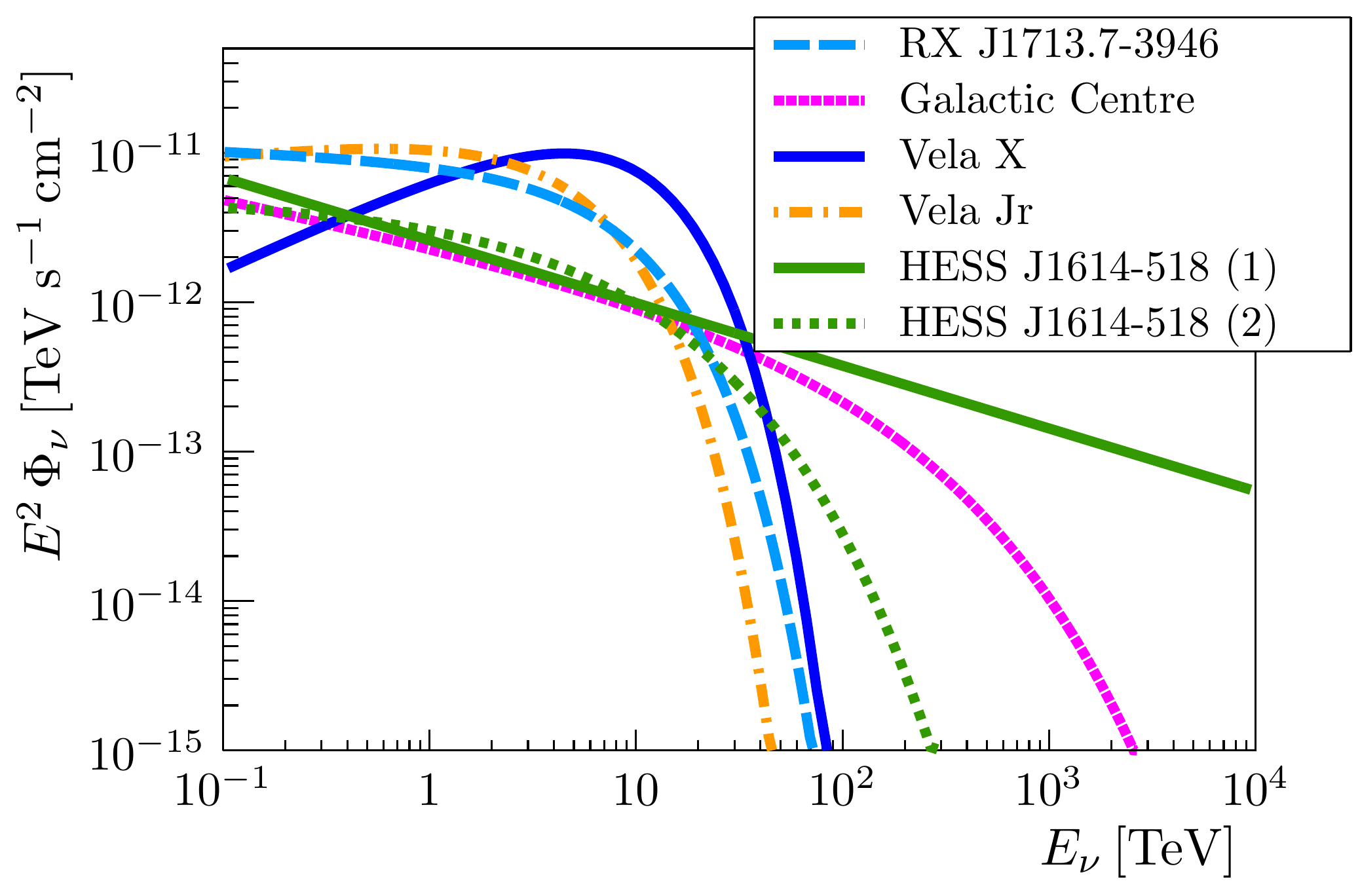}\
\centering
\includegraphics[width=0.6\textwidth]{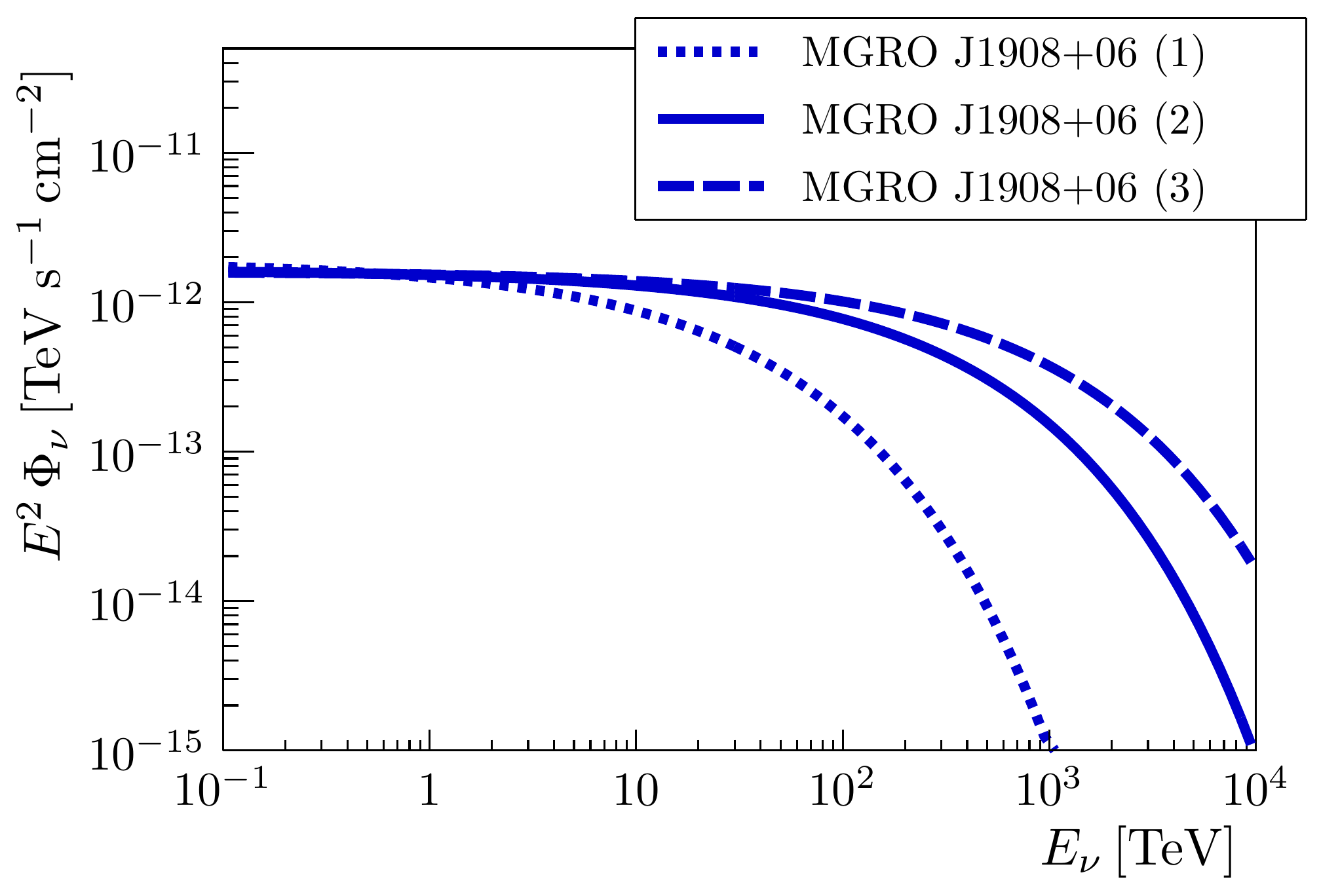}
\caption{Muon neutrino fluxes ($\nu_\mu + \bar{\nu}_{\mu}$) used in the analysis. The corresponding parameters are given in Table~\ref{tab:source}.}
\label{fig:flux}
\end{figure}

All neutrino fluxes in this publication are parameterised by
\begin{equation}
 \Phi_\nu(E) =k_0 
  \left(\frac{E}{1\,\text{TeV}}\right)^{-\Gamma} 
           \exp\left[-\left(\frac{E}{E_\text{cut}}\right)^\beta\right],
\label{eq:flux}
\end{equation}
\noindent where $k_0$ is the normalisation constant, $\Gamma$ is the spectral index, $E_\text{cut}$ is the energy cutoff, $\beta$ is the cutoff exponent \cite{RXJ1713_HESS}.
Table~\ref{tab:source} lists the sources considered, their
declination $\delta$ and angular extension (indicated as radius), as measured by $\gamma$-ray detectors, as well as the parameters of the Eq.\ (\ref{eq:flux}). For several sources,
different parameterisations are consistent with the $\gamma$-ray data
and the corresponding neutrino fluxes are included in the analysis. The fluxes listed in Table~\ref{tab:source} 
are shown in Fig.~\ref{fig:flux}.

\begin{table}[ht]
\centering
\caption{Parameters of the candidate sources investigated, references for the corresponding $\gamma$-ray measurements and source type. The neutrino flux is expressed according to Eq.\~(\ref{eq:flux}), with the normalisation constant $k_{0}$ in units of $10^{-11}  \, \text{TeV}^{-1}  \, \text{s}^{-1}  \, \text{cm}^{-2}$ and $E_\text{cut}$ in units of TeV. See the text for further details (note that $\xi_\text{had} = 1$ is assumed).}
\vspace{10pt}
\label{tab:source}
\resizebox{\textwidth}{!}{\begin{tabular}{lllllllll}  

\hline
Source				&	$\delta$		&	radius		&	$k_{0}$	&	$\Gamma$	&	$E_\text{cut}$	&	$\beta$  & $\gamma$-ray data  & type\\	\hline
\hline

RX\,J1713.7-3946		&	-39.77$^\circ$	&	0.6$^\circ$	&	0.89		&	2.06			&	8.04		&	1	&	 \cite{HESS16}	& SNR\\		\hline

Vela~X				&	-45.6$^\circ$	&	0.8$^\circ$	&	0.72		&	1.36			&	7		&	1 	&	 \cite{velaXnew}& PWN		\\		\hline

Vela Jr				&	-46.36$^\circ$	&	1$^\circ$		&	1.30		&	1.87			&	4.5		&	1	&	 \cite{VelaJrHESS2016}& SNR		\\		\hline

HESS\,J1614-518 (1)		&	-51.82$^\circ$	&	0.42$^\circ$	&	0.26		&	2.42			&	-		&	-	&	\cite{HESSJ1614} & SNR\\		\hline

HESS\,J1614-518 (2)		&	-51.82$^\circ$	&	0.42$^\circ$	&	0.51		&	2			&	3.71		&	0.5	&	\cite{HESSJ1614} & SNR\\		\hline

Galactic Centre			&	-28.87$^\circ$	&	0.45$^\circ$	&	0.25		&	2.3			&	85.53	&	0.5	&	 \cite{Nature}& UNID	\\ \hline

MGRO\,J1908+06 (1) 	&	6.27$^\circ$	&	0.34$^\circ$	&	0.18		&	2			&	17.7	&	0.5	&	 see text	& UNID\\ \hline

MGRO\,J1908+06 (2)		&	6.27$^\circ$	&	0.34$^\circ$	&	0.16		&	2			&	177	&	0.5	&	 see text	& UNID\\ \hline

MGRO\,J1908+06 (3)		&	6.27$^\circ$	&	0.34$^\circ$	&	0.16		&	2			&	472	&	0.5	&	 see text	& UNID\\ \hline

\hline
\end{tabular}
}
\end{table}

In the following subsections short descriptions of the sources are given with details on
the derivation of the neutrino flux from the measured $\gamma$-ray flux.

\subsection{RX\,J1713.7-3946}

The young shell-type SNR RX\,J1713.7-3946 is at
present one of the best studied SNRs in the VHE regime. Its high-energy $\gamma$-ray emission has
been observed by H.E.S.S.\ in several campaigns, in the years 2003-2005
\cite{HESS04, HESS06, RXJ1713_HESS} and in 2011 and 2012 \cite{HESS16}. The
reported spectrum extends up to about 100\,TeV, suggesting that the hadronic
particle population may have energies up to several PeV if the $\gamma$-ray production
is hadronic. 

The origin of the TeV $\gamma$-ray emission from RX\,J1713.7-3946 has been a
matter of active debate. A detailed discussion of the interpretation of the H.E.S.S.\
data in hadronic or leptonic scenarios can be found e.g.\ in Refs.\ \cite{HESS16, HESS06} and references therein. 
Fermi-LAT reported an observation of GeV $\gamma$-ray emission from RX\,J1713.7-3946 \cite{Fermi_RXJ1713}.
While the hard spectrum at GeV energies reported by the Fermi Collaboration is
generally interpreted as an argument in favour of a leptonic scenario, some
authors argue that both hadronic and leptonic scenarios can
reproduce the data under certain assumptions (e.g.\ Ref.\ \cite{Gabici}). 
On the other hand, the observation of molecular clouds in the vicinity of the source \cite{Fukui, Maxted} could provide an additional hint in favour of the hadronic scenario. 
In Ref.\ \cite{Silvia}, a detailed numerical treatment of the SNR shock interaction in a non homogenous medium has been reported, consistently describing the broadband GeV--TeV spectrum of RX\,J1713.7-3946 in terms of a hadronic model.
In Ref.\ \cite{HESS16}, the X-ray, the Fermi-LAT and the
updated H.E.S.S.\ data are combined to derive in both scenarios the particle
spectra from the SNR spectral energy distribution. The data can be fit both
with hadronic and leptonic models so neither of the two scenarios can currently
be excluded.

Previous KM3NeT results \cite{loi} were derived from H.E.S.S. data in Ref.\ \cite{RXJ1713_HESS}. These are superseded by a most recent H.E.S.S. publication \cite{HESS16} which shows a softer spectrum at the highest energies compared to the previous paper. 
This new spectrum is based on a new analysis which makes use of more data, refined calibration and data analysis methods. 
The results presented in this publication are based on the data reported in Ref.\ \cite{HESS16}.
In Table~\ref{tab:source} only the flux derived with the method in Refs. \cite{Vissani, Vissani_RXJ1713, Vissani_formule} is reported.

\subsection{Vela~X}

Vela~X is one of the nearest pulsar wind nebulae and is associated with the
energetic Vela pulsar PSR\,B0833-45. Even if PWNe are generally considered as
leptonic sources, interpretation of TeV $\gamma$-ray emission from Vela~X in
terms of hadronic interactions has been discussed \cite{horns,VelaHadr2}.

The VHE $\gamma$-ray emission from Vela~X was first reported by the H.E.S.S.\
Collaboration \cite{velaXold} and was found to be coincident with a region of
X-ray emission discovered with ROSAT as a filamentary structure extending
south-west from the pulsar to the centre of Vela~X. 
The first result of H.E.S.S. has been updated \cite{velaXnew} with data from the
2005--2007 and 2008--2009 observation campaigns and using an improved method
for the background subtraction. 
The new data are characterised by a 25\% higher integral flux above 1 TeV and a harder energy spectrum and are used here to derive the neutrino spectrum.

\subsection{Vela Jr}

RX\,J0852.0-4622, commonly referred as Vela Junior (Vela Jr in
Table~\ref{tab:source} and Fig.~\ref{fig:flux}) is a young shell-type SNR with
properties similar to RX\,J1713.7-3946. Vela Junior emits $\gamma$-rays up to energies of few tens of TeV \cite{VelaJrHESS2016}. 
Also for this source the $\gamma$-ray emission has been interpreted
both in the hadronic and leptonic scenarios (see Ref.\ \cite{VelaJrHESS2016} for an
overview on the arguments). In particular, a recent analysis \cite{VelaJrFukui}
reports a good spatial correspondence between the TeV $\gamma$-rays and
interstellar hydrogen clouds, suggesting a hadronic interpretation of the origin of the observed $\gamma$-rays from this source.

\subsection{HESS\,J1614-518}

The $\gamma$-ray high energy emission of the source HESS\,J1614-518 has been
observed by H.E.S.S. up to about 10 TeV \cite{HESSJ1614} and was studied in terms of morphological, spectral and multi-wavelength
properties and classified as candidate shell-type SNR. 
The $\gamma$-ray
flux from the source HESS\,J1614-518 has been fitted in Ref.\ \cite{HESSJ1614} as a
pure power law and the neutrino flux derived from it is indicated in the
following as HESS\,J1614-518 (1). To test the effect of a possible cutoff in the
spectrum, in this study the H.E.S.S. $\gamma$-ray data were fitted also with a power law
with exponential cutoff. The neutrino flux derived from this 
$\gamma$-ray flux is referred as HESS\,J1614-518 (2).

\subsection{Galactic Centre}

Recently, the H.E.S.S.\ Collaboration has reported $\gamma$-ray
observations of the region surrounding the Galactic Centre \cite{Nature}. The
$\gamma$-ray flux reported is derived for two regions: a point source with radius 0.1$^\circ$ (PS) HESS\,J1745-290 centred on Sgr A* and a diffuse emission (DF)
from an annulus between 0.15$^\circ$ and 0.45$^\circ$. The strong correlation
between the brightness distribution of diffuse VHE $\gamma$-ray emission in the
wider vicinity of the Galactic Centre and the locations of molecular clouds
points towards a hadronic origin of the $\gamma$-ray emission \cite{Nature}. 
Since the DF $\gamma$-ray data are consistent with a hard power-law, the spectrum of the parent protons should extend to PeV energies. 
The neutrino spectra expected from the two regions PS and DF have been evaluated
in Ref.\ \cite{VissaniGC}, where a few possible neutrino spectra are proposed starting from plausible $\gamma$-ray fluxes and exploring different energy cutoffs. The flux considered here is the sum of the PS and DF
regions, choosing as flux for the PS area the one derived from the $\gamma$-ray flux with $E_{\text{cut},\gamma}=10.7$ TeV and for the DF region the one from $E_{\text{cut},\gamma}=0.6$ PeV. 
For simplicity, the source shape is approximated as a homogeneous disk of radius 0.45$^\circ$.

\subsection{MGRO\,J1908+06}

The source MGRO\,J1908+06 has been detected both by air Cherenkov telescopes
(H.E.S.S.\ \cite{HESSMGRO} and VERITAS \cite{VERITASMGRO}) and extensive
air-shower detectors (Milagro \cite{MGRO1, MGRO2, MGRO3}, ARGO-YJB
\cite{ARGOMGRO} and HAWC \cite{HAWCMGRO}). The nature of this source is currently unclear. It could be a
PWN associated with the pulsar PSR J1907+0602 \cite{MGROPWN}. Its large size and the lack of
softening of the TeV spectrum with distance from the pulsar, however, are
uncommon for TeV PWNe of similar age, suggesting that it could also be a SNR  \cite{VERITASMGRO, HalzenMGRO}. Using the measured $\gamma$-ray spectra, the prospects for
detecting neutrinos from this source with IceCube are discussed in Ref.\
\cite{HalzenMGRO}. Three possible assumptions on the $\gamma$-ray flux are
considered, with a spectral index $\Gamma_\gamma=2$ and cutoff energies
$E_{\text{cut},\gamma}=30, 300, 800$\,TeV. The corresponding neutrino fluxes
derived in Ref.\ \cite{HalzenMGRO}, listed in Table~\ref{tab:source} as (1), (2) and
(3), respectively, are used in this analysis. The source position and extension
are taken from the H.E.S.S.\ results \cite{HESSMGRO}.

\section{Simulations}
\label{sec:simul}

For this analysis the Monte Carlo (MC) chain discussed in Ref.\ \cite{loi} is used.
Neutrinos and anti-neutrinos of all flavours are considered, and both charged and
neutral current reactions are simulated. 
For the generation equal fluxes of neutrinos and anti-neutrinos are assumed.
Since neutrino and anti-neutrino
interactions cannot be distinguished in KM3NeT on an event-by-event basis, the lepton
symbols ($\nu$, $\mu$, $e$, and $\tau$) denote both particles and anti-particles
in the following. 
Neutrinos are generated over the full solid angle to simulate
the background of atmospheric neutrinos. Neutrinos from the specific sources
described in Section~\ref{sec:nuflux} are simulated as originating from
homogeneous disks centred at the declination shown in Table~\ref{tab:source} and
with a radius given in the same table. 
Events are generated in the energy range between $10^2$ and $10^8$\,GeV
according to an $E^{-1.4}$ spectrum and subsequently reweighted to different flux
models. The neutrino interactions are simulated using LEPTO \cite{lepto} with the parton
distribution functions CTEQ6 \cite{CTEQ} (for deep inelastic scattering). The
muon produced at the interaction vertex is propagated through rock and water
with MUSIC \cite{music}.

The Cherenkov photons induced by charged particles traversing the water are
propagated to the DOMs. To save CPU time, this is done using tabulated photon
propagation probabilities based on full GEANT3.21 \cite{GEANT} simulations and taking into
account the DOM properties (effective area, quantum efficiency and
collection efficiency of the photomultipliers; transmission probability through
glass and gel), the DOM orientation with respect to the incident direction of the photon and the optical water properties measured at the KM3NeT-Italy site
\cite{giorgio}. For each event, the PMTs measuring a signal are determined, each
signal (``hit'') being characterised by the photon arrival time and the signal
amplitude (deposited charge). The hit
data are converted to digitised arrival time and time-over-threshold (ToT),
i.e.\ the time the analog signal exceeds a predefined threshold.

Optical background due to the presence of $^{40}$K in salt water is simulated by
adding an uncorrelated hit rate of 5\,kHz per PMT. Moreover, the probability of
two-, three- and four-fold hit coincidences on a DOM from a single $^{40}$K
decay have been estimated by GEANT simulations and are included with rates of
500, 50 and 5\,Hz per DOM, respectively. Both the single and coincidence rates
are in agreement with the results from the prototype detection unit of the
KM3NeT detector deployed at Capo Passero \cite{DOM}. The effect of
bioluminescence light is negligible at the KM3NeT-Italy site \cite{pellegriti}.

At the end of the simulation chain, trigger algorithms are applied in order to
select potentially interesting events that will be reconstructed and analysed
with the statistical methods described below. The trigger is based on the L1 hits, i.e.\ hits on more than one PMT of the same DOM in a time
window of 10\,ns. Events pass the trigger condition if there are at least 5
causally connected L1 hits. Details on the trigger and trigger efficiency are given in Ref. \cite{loi}.

\subsection{Atmospheric neutrinos and muons} 

Only a very small fraction of the high energy neutrino flux arriving at the detector is
of astrophysical origin. The dominant contribution is due to atmospheric
neutrinos from extended air showers caused by cosmic ray interactions with
 nuclei in the atmosphere. However, at sufficiently large energies, the astrophysical flux will dominate that of atmospheric origin.
 The atmospheric neutrino flux has two
components: the conventional one due to the decay of charged pions and kaons and
the prompt one due to the decay of charmed hadrons, produced in the
primary interaction. The atmospheric neutrino flux is simulated assuming the
conventional atmospheric model as in Ref.\ \cite{Honda} and the prompt
component as described in Ref.\ \cite{Enberg}. Corrections due to the
break in the cosmic ray spectrum (knee) are applied as described in Ref.\ \cite{knee}. 
Also other models of prompt neutrino fluxes \cite{prompt1,prompt2,prompt3} have been tested, but they leave the final results essentially unaltered.

In addition to atmospheric neutrinos, cosmic ray interactions in the atmosphere also produce atmospheric muons. 
Each initial interaction creates a number of muons that are collimated and coincident in time (muon bundle). 
Atmospheric muons are simulated using the MUPAGE event generator \cite{mupage}. 
In the analysis presented here two simulated muon event samples are used, one with muon bundle energies $E_\text{b}>10\,$TeV, corresponding to a livetime of about 3~months, the other with $E_\text{b}>50$\,TeV, equivalent to about $3$ years of livetime.

\section{Event reconstruction performances}
\label{sec:reco}

The neutrino induced events are observed in two topologies, track-like and
cascade-like events, each class requiring specific event reconstruction
algorithms.

Track-like events are due to charged-current $\nu_\mu$ interactions that, for
$E_\nu\gtrsim 1$\,TeV, produce in the final state muons with track lengths of
the order of kilometres and trajectories almost colinear with the parent
neutrino direction. Also $\nu_\tau$ charged-current interactions can produce a
high-energy muon in the final state through a muonic decay of the final-state
$\tau$ with a branching ratio of about 17\%. The reconstruction algorithm used for track-like
events is described in Ref.\ \cite{Karel}. The muon direction is reconstructed
from the sequence of Cherenkov photon hits on the PMTs, taking advantage of the
fact that photons are emitted along the particle track at an angle of about $
42^\circ$. 
The reconstruction algorithm starts by a prefit scanning the full solid angle. Then, starting from the twelve best fitted directions in the prefit, a maximum
likelihood search is performed. The likelihood is derived from a probability density function depending
on the position and orientation of the PMTs with respect to the muon trajectory
and on the hit times.
Among these intermediate tracks, the one with the best likelihood is chosen. 
A reconstruction quality parameter is defined as $\Lambda = - \log L - 0.1N_\text{comp}$, where $\log L$ is the log-likelihood of the fit and $N_\text{comp}$ is the number of intermediate tracks during the reconstruction within $1^\circ$ from the chosen one.

The angular resolution, calculated as the median angle
between the reconstructed track and the neutrino direction, is smaller than $0.2^\circ$
at $E_\nu >10$\,TeV. The energy is reconstructed from the spatial distribution
of hit and non-hit PMTs. The resolution is better than 0.3 units in
$\log_{10}(E_\text{reco}/E_\mu)$, where $E_\text{reco}$ is the reconstructed and
$E_\mu$ the true muon energy at the detector level.

All neutral-current reactions, as well as charged-current reactions of $\nu_e$
and most $\nu_\tau$, produce cascade-like event topologies. Particle cascades
evolve from the hadronic final state and the final-state charged lepton ($e$) or
its decay products ($\tau$, except in muonic decays). These cascades are typically several
metres long and therefore small compared to inter-DOM distances. 
The reconstruction for such cascade-like events has an angular resolution worse than the track-like case and is described in \cite{Karel}.

Track-like events are of particular relevance for the search for point-like,
i.e.\ very localised sources of neutrino emission, since they allow for fully
exploiting the large effective area and the good angular resolution of
KM3NeT/ARCA. The analysis discussed in this paper therefore focusses on track-like
events, and consequently the event reconstruction specific for track-like events
is applied to all events, including cascade events. 
Only events with sufficient reconstruction quality are retained.

\section{Galactic sources: Search method and results}
\label{sec:analysis}

Since the neutrino signal from any point source must be identified on top of a
large background of atmospheric muons and neutrinos, statistical techniques are
required to quantify a possible excess of events around the source position. The
two quantities used to describe the detector performance are the discovery
potential and the sensitivity. 
The discovery potential refers to the flux that could produce a significant (e.g.\ 3$\sigma$ or 5$\sigma$) observation with probability 50\%.
The sensitivity refers to the flux that can be excluded at a given confidence level (90\% in this paper), if no significant signal is observed (see Section~\ref{sec:unbinned}).

The search for Galactic point-like neutrino sources is performed in the
following steps:
\begin{itemize}
\item 
Selection cuts are applied to reduce the background events (Section~\ref{sec:cuts}).

\item 
A multivariate analysis employing a Random Decision Forest algorithm \cite{RFprimo} is performed on the remaining events to distinguish signal from background events (Section~\ref{sec:RF}).

\item 
An unbinned likelihood method is used to determine the discovery potential and
the sensitivity (Section~\ref{sec:unbinned}).
\end{itemize}

\subsection{Selection cuts}
\label{sec:cuts} 

For signal events, only charged-current interactions of $\nu_\mu$ and of
$\nu_\tau$ (with subsequent $\tau\to\mu\nu\overline{\nu}$ decay) are considered since the
remaining event classes (other decays of $\nu_\tau$ producing cascades, charged-current $\nu_e$ and all neutral-current interactions) are almost completely rejected by applying track reconstruction quality criteria. 
For atmospheric neutrinos, both charged- and neutral-current
$\nu_\mu$ and $\nu_e$ events are taken into account.

The loose selection cuts applied are:
\begin{enumerate}
	\item A zenith cut at about 10$^\circ$ above the horizon to reduce the background of atmospheric muons (see Section~\ref{sec:intro}), slightly optimised for each candidate source taking into account its maximum elevation (see Table~\ref{tab:events}).
	\item A cut on the angle $\alpha$ between the reconstructed track direction and the nominal source position. A cut $\alpha < 10^\circ$ has been selected as a compromise to reduce the background without reducing significantly the efficiency for selecting signal events. 
\end{enumerate}
The numbers of signal events after these selection cuts, expected from the different sources for the flux assumptions from Table~\ref{tab:source}, are reported in Table~\ref{tab:events}. 

\begin{table}[h]
\centering
\caption{
Zenith cut ($\theta_\text{cut}$) and expected number of signal events for the
candidate sources in five years of data taking. The number of events is specified at three stages: after
reconstruction; after zenith cut; and after the $\alpha$ cut (see text). The sum of
$\nu_\mu$ and $\nu_\tau$ events is shown, where the $\nu_\tau$ contribution is
between 8\% and 10\%. }
\label{tab:events}
\begin{tabular}{lcccc}
\hline
Sources 	  		& 	 $\theta_\text{cut}$	& Reconstructed 	&	Events with 				&	Events with \\	
				&	 [$^\circ$]					& events			&	$\theta \geqslant \theta_\text{cut}$	&	$\theta \geqslant \theta_\text{cut}$ \\	
				&						&				&							&	AND $\alpha \leqslant 10^\circ$ \\	\hline

\hline				

RX\,J1713.7-3946	& 	 78					& 	 22.0			& 	20.0						& 	  16.4\\	\hline

Vela~X 			& 	 81	  				& 	 41.5	 		& 	40.7	 					& 	  34.9\\	\hline

Vela Jr  	  		& 	 80	  				& 	 26.0	 		& 	25.6	 					& 	  21.1\\	\hline

HESS\,J1614-518 (1)	& 	 86	  				& 	 10.7	 		& 	10.5	 					& 	  9.1\\	\hline

HESS\,J1614-518 (2)	& 	 86	  				& 	 9.3	 		& 	9.1	 					& 	  7.7\\	\hline

Galactic center  	& 	 78	 				 & 	 9.1	 		& 	7.0	 					& 	  5.7\\	\hline

MGRO\,J1908+06 (1)	& 	 80	 				 & 	 6.7	 		& 	4.1	 					&	   3.5\\	\hline

MGRO\,J1908+06 (2)	& 	 80	 				 & 	 11.9	 		& 	7.1	 					&	   6.1\\	\hline

MGRO\,J1908+06 (3)	& 	 80	 				 & 	 14.0	 		& 	8.3	 					&	   7.1\\	\hline

\hline
\end{tabular}
\end{table}

\subsection{Random Decision Forest training}
\label{sec:RF}
A multivariate analysis employing the Random Decision Forest algorithm is
performed to distinguish three classes of events: neutrinos coming from the
source, atmospheric neutrinos, and atmospheric muons.
More specifically we use the extremely randomised trees classifier from Ref. \cite{scikit-learn}.

\begin{figure}[h]
	\begin{subfigure}{.5\textwidth}
  		\includegraphics[width=0.9\linewidth]{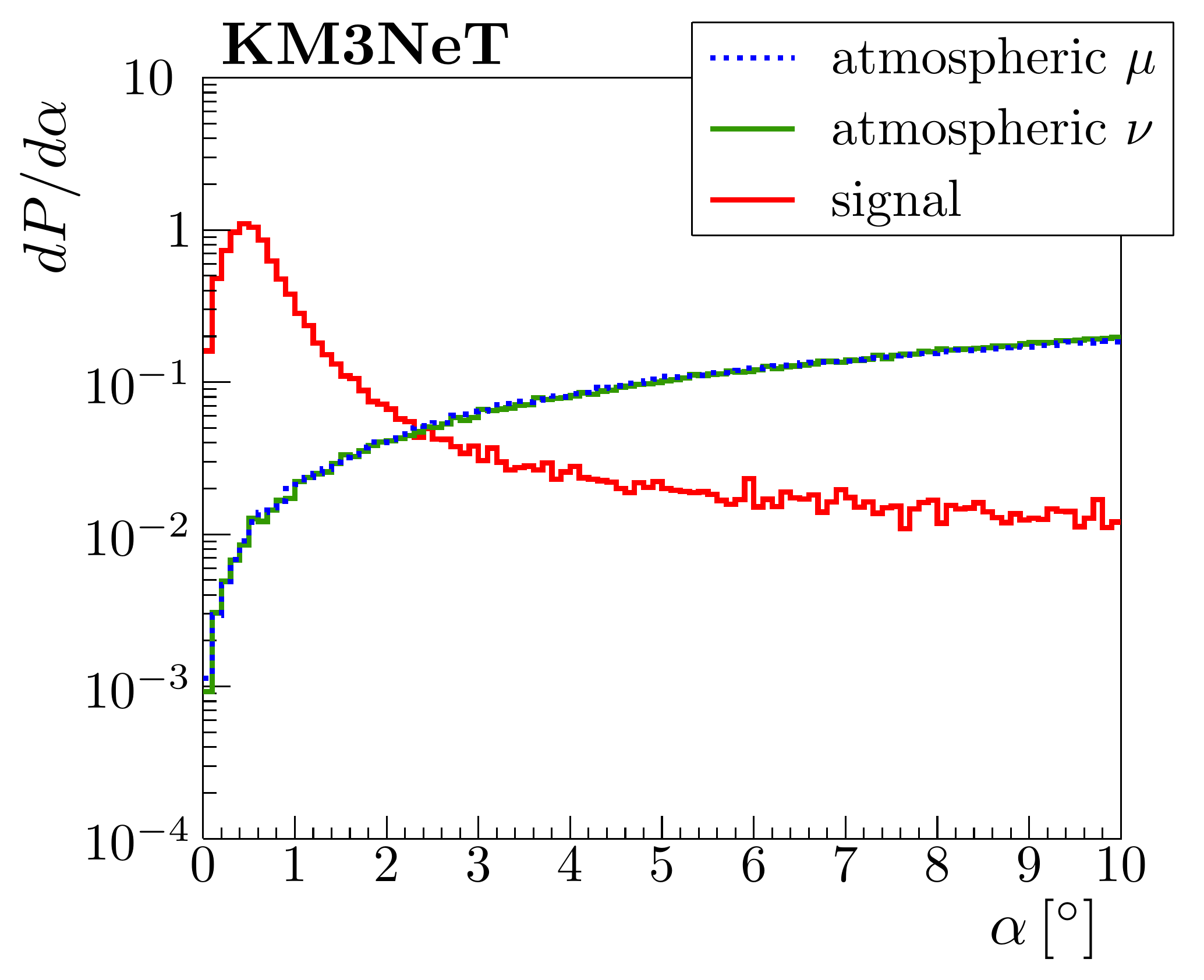}
	\end{subfigure}%
	\begin{subfigure}{.5\textwidth}
  		\includegraphics[width=0.9\linewidth]{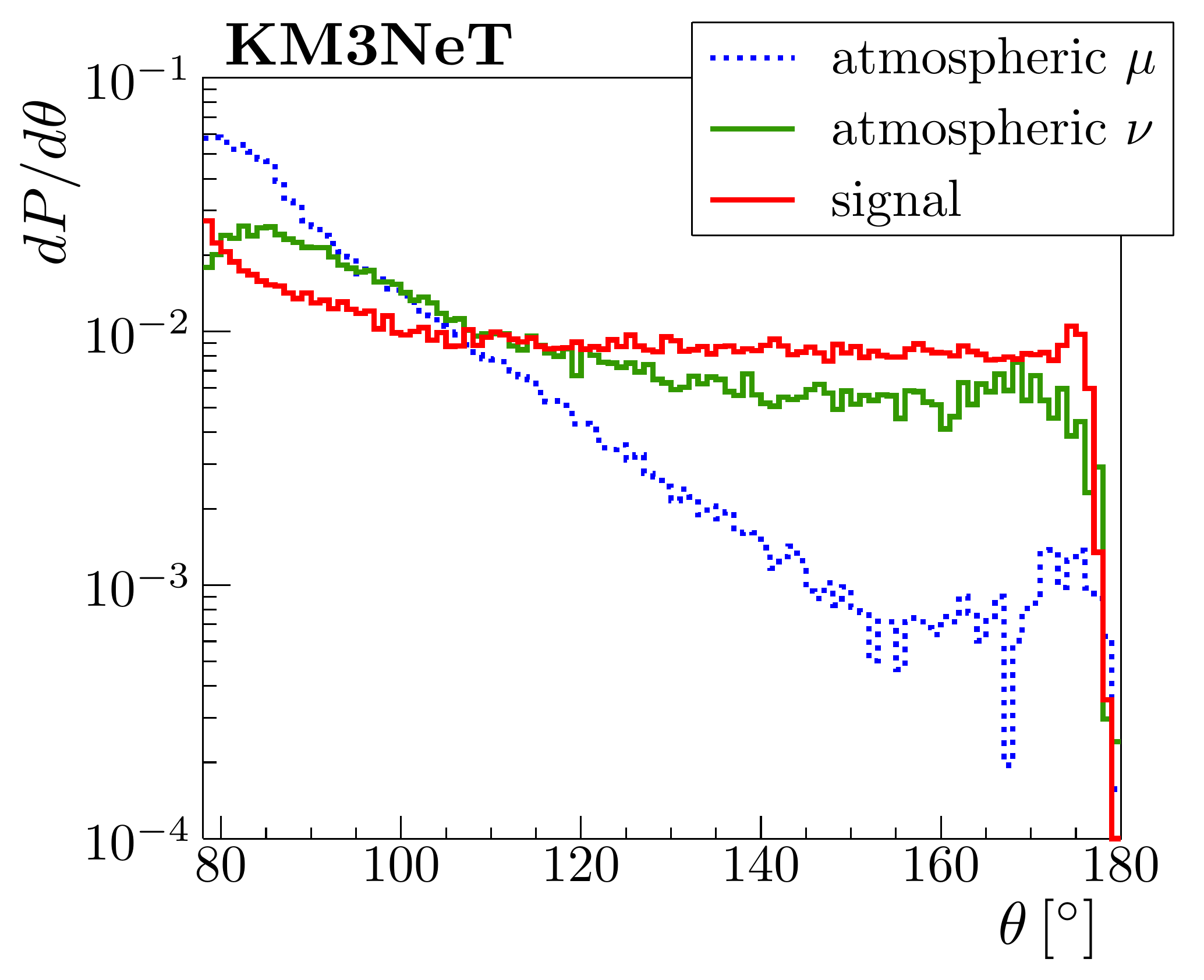}
	\end{subfigure}
	\begin{subfigure}{.5\textwidth}
  		\includegraphics[width=0.9\linewidth]{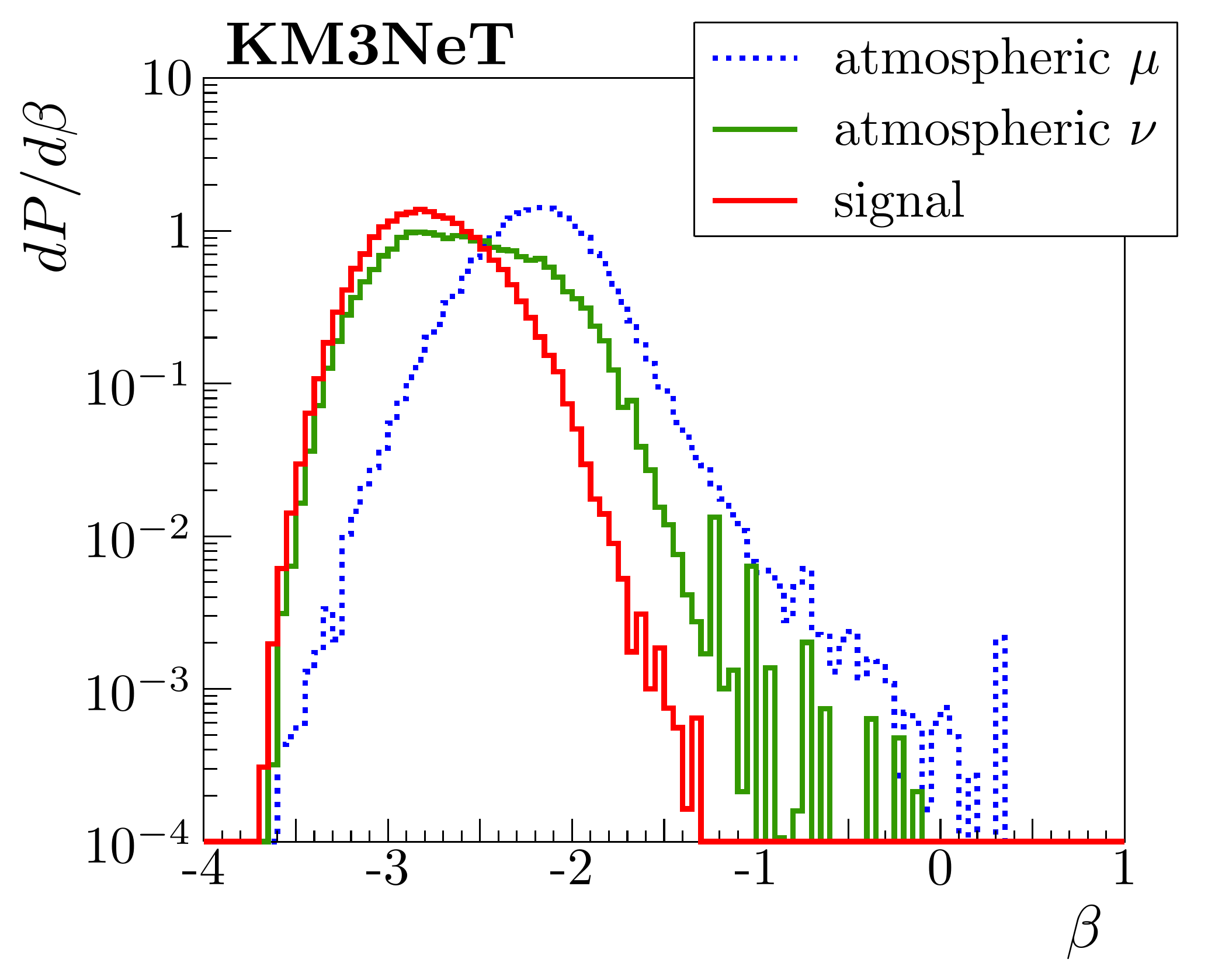}
	\end{subfigure}%
	\begin{subfigure}{.5\textwidth}
  		\includegraphics[width=0.9\linewidth]{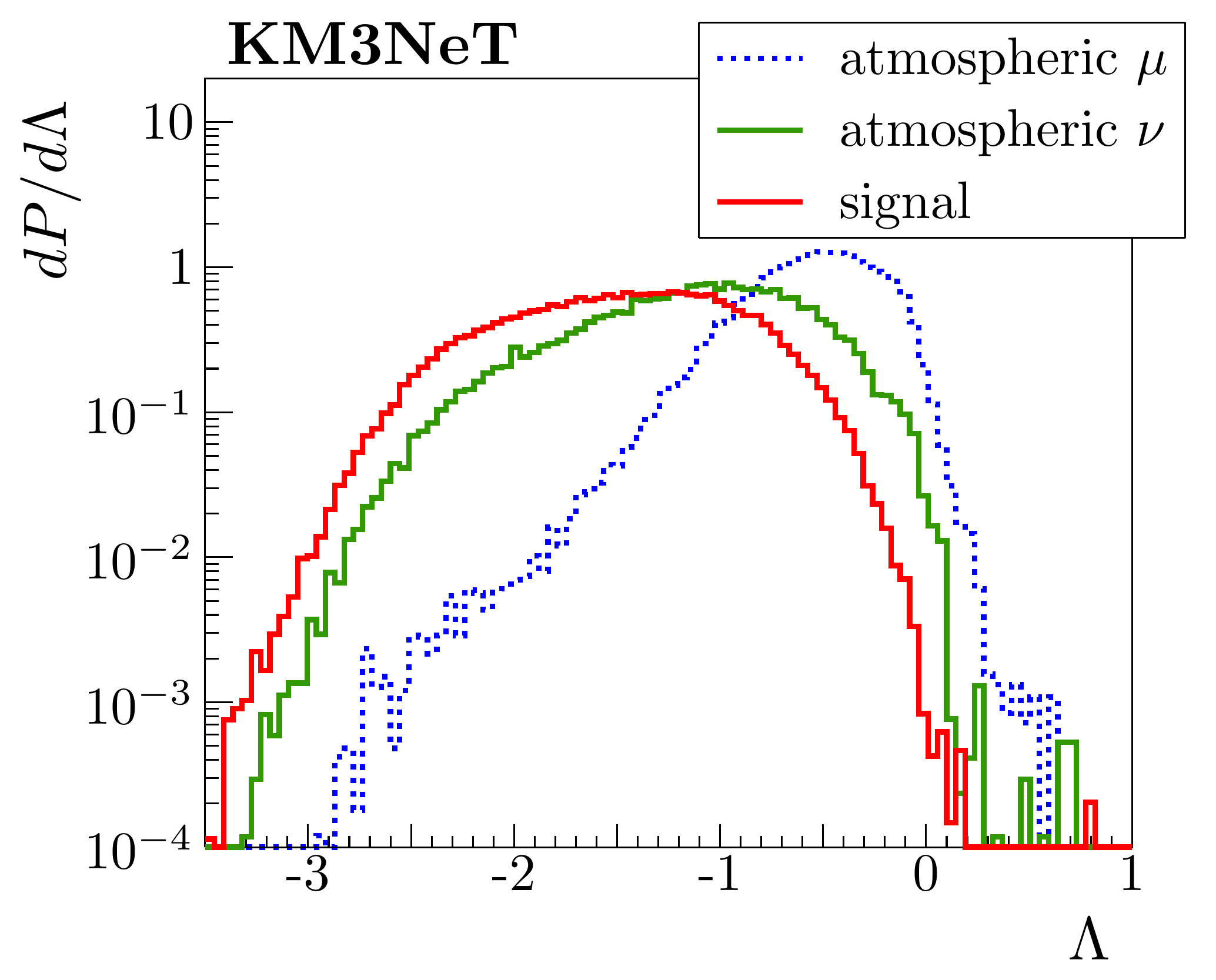}
	\end{subfigure}
\caption{
Distributions of the most significant features used in the training ($\alpha, \theta, \beta, \Lambda$), for the
three different event classes: atmospheric muons (blue lines), atmospheric
neutrinos (green lines) and neutrinos from the source (red lines). Shown are the
distributions for SNR RX\,J1713.7-3946 (see Table~\ref{tab:source}) with the
zenith and $\alpha$ cuts applied. In the top left plot the blue and green lines are superimposed. }
\label{fig:figtest}
\end{figure}

The features used in the training to characterise the events are: 
the angle $\alpha$ between the reconstructed track direction and the nominal source position; 
the reconstructed zenith angle $\theta$; 
the reconstructed muon energy at the detector level; 
the numbers of hits used at different stages of the reconstruction; 
the error estimate on this fit $\beta$; 
and the track reconstruction quality parameter, $\Lambda$, defined in Section \ref{sec:reco}.
 The distributions of the most important of these features are shown
in Fig.~\ref{fig:figtest} for the three event classes. 
Note that $\alpha$ is the convolution of the source extension and the angular resolution.

\begin{figure}[h]
\begin{center}
\includegraphics[width=0.7\textwidth]{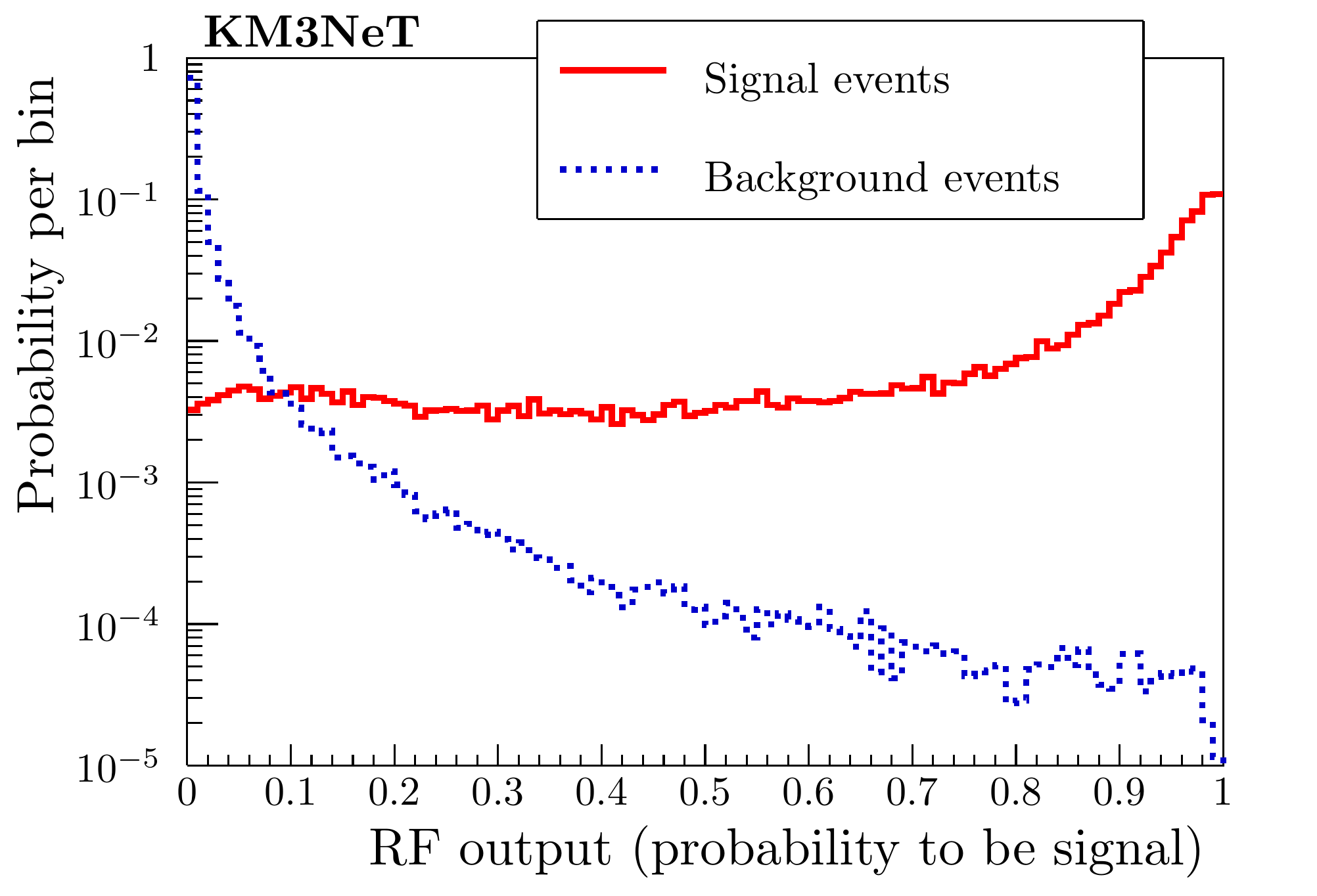}
\end{center}
\caption{
Distribution of the probability to be signal according to the Random Decision Forest output for signal and backgrounds events for
the source RX\,J1713.7-3946.}
\label{fig_PDF}
\end{figure}

In a first step the algorithm is trained on a sample of events to optimise its
performance in distinguishing the different event classes. The trained classifier
has then been applied to a separate event sample to test its performance. For each event the classifier returns the probability to belong to each one of the three classes. 
The distributions of the probability to belong to the signal class (Fig. \ref{fig_PDF}) are used for all events as probability density functions in the subsequent analysis step.

\begin{figure} [h]
\begin{center}
\includegraphics[width=0.7\textwidth]{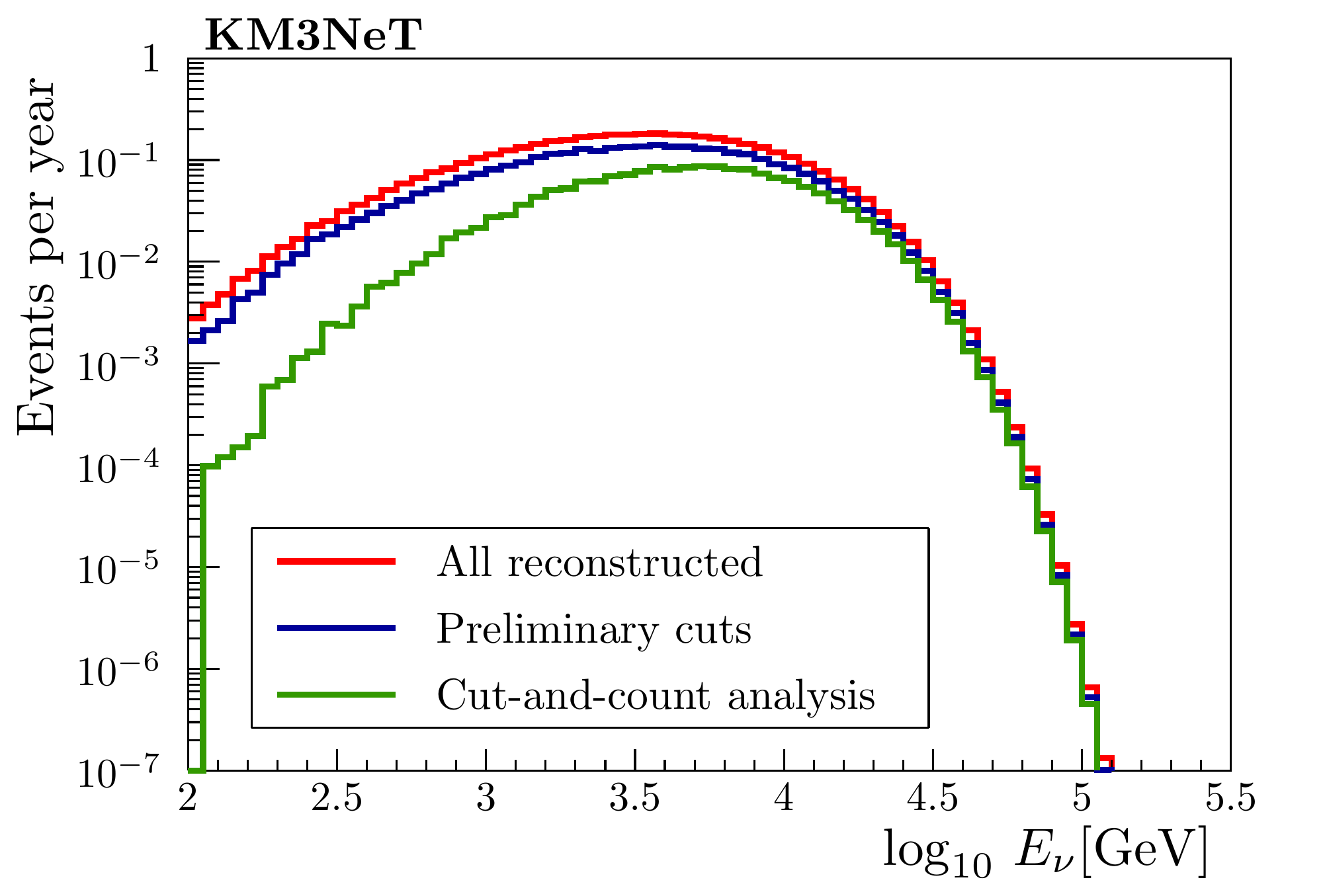}
\end{center}
\caption{
Distribution of the generated neutrino energy for the source RX\,J1713.7-3946 at reconstruction level (red line), after the selection cuts $\theta \geqslant 78 ^\circ$ and $\alpha < 10^\circ$ (blue line) and after the cuts of the cut-and-count analysis (green line), that corresponds to an additional cut on the Random Decision Forest output greater than 0.92.}
\label{fig:E}
\end{figure}

An example of the distribution of the simulated neutrino energy at the different stages of the analysis is shown in Fig.~\ref{fig:E} for the source SNR RX\,J1713.7-3946 (see Table~\ref{tab:source}).
The energy distribution for all the reconstructed events, those passing the selection cuts (see Table~\ref{tab:events}) and those passing the cuts of the ``cut-and-count'' analysis (see \cite{bubbles} for a description of this method) is shown. In latter case, the output of the Random Decision Forest classifier is used a as variable to cut on. This is shown here to illustrate the energy of interest of these kind of analyses, typically peaking around 10 TeV. 
However, the cut-and-count method is not used for the results described in the next sections.

\subsection{Unbinned method}
\label{sec:unbinned}

In order to test the compatibility of the data with two different hypotheses
$H_0$ and $H_1$, a test statistic is defined. The test
statistic can in principle be any function of the data but is optimally selected such that its distributions under the two competing hypotheses are maximally separated.
In the search for neutrino point sources the hypothesis $H_0=H_\text{b}$ refers
to the case in which the data set consists of background events only. Hypothesis $H_1 = H_\text{s+b}$ refers to the case where events from a cosmic source
are present in addition to the background. To calculate the test statistic, a likelihood ratio \cite{Neyman}
has been defined as the ratio of the probabilities to obtain the data assuming the hypotheses $H_\text{s+b}$ or $H_\text{b}$:
\begin{equation}
\text{LR} = \log \left[ \frac{P(\text{data}|H_\text{s+b})}
                           {P(\text{data}|H_\text{b})} \right]\,.
\end{equation}

The likelihood ratio can be written in terms of the probability density
functions (PDFs) describing the distribution of signal and background events as
a function of a given variable $x$, $\text{f}(x|\text{s})$ and $\text{f}(x|\text{b})$:
\begin{equation}
\text{LR}=\sum_{i=1}^n\log\left[
          \frac{\dfrac{n_\text{s}}{n}\cdot\text{f}(x_{i}|\text{s}) + 
                \left(1-\dfrac{n_\text{s}}{n}\right)\cdot\text{f}(x_{i}|\text{b})}
               {\text{f}(x_{i}|\text{b})}\right]\,,
\label{eq_cap5_15}
\end{equation}

where $n$ is the total number of recorded events in a given period of time and $n_\text{s}$ is the expected number of signal events in the sample of $n$ events; $n_\text{s}$ is a free parameter constrained to be non-negative. Note that the source position is assumed to be known and is not determined from the data. For each sample, $\text{LR}$ is maximised.
The maximum value of $\text{LR}$ is used as the test statistic and will be denoted with the symbol $\lambda$. The variable
$x$ in Eq.~(\ref{eq_cap5_15}) is the probability that the event belongs to the
``signal'' class as calculated by the Random Decision Forest classifier. As an example the PDFs for the source RX\,J1713.7-3946 are shown in Fig.~\ref{fig_PDF}.

The output of the algorithm is the $\lambda$ value and the
corresponding fitted $n_\text{s}$ value. 
The distribution of $2\lambda$ for the background-only case is expected to follow a half-$\chi^2$-distribution as defined in Ref. \cite{cowan} and can be used to estimate the pre-trial $p$-value.

In order to estimate the distribution of the test statistic for the background-only assumption, the algorithm is applied to several thousand samples of background events sampled
from the simulated atmospheric neutrino and muon events. 
For each sample, the maximum value of LR, $\lambda$, is recorded. 
The normalised distribution of $\lambda$, g($\lambda|$b) is then determined. 
Selecting the required significance and the corresponding two-sided Gaussian probability, e.g.\ $3\sigma$ and $2.7 \times 10^{-3}$, a critical value $\lambda_{3\sigma}$ is calculated from

\begin{equation}
 \int_{\lambda_{3\sigma}}^\infty g(\lambda | \text{b}) \text{d}\lambda
 = 2.7 \times 10^{-3}\,.
\label{critical}
\end{equation}

Subsequently, the procedure is repeated adding the poissonian expectation, $N_\text{s}$, of one simulated signal event to the background sample, then the poissonian expectation for two signal events, and so on.
For each $N_\text{s}$, $\lambda$ is again calculated and its normalised distribution will be indicated with g($\lambda | N_\text{s} + \text{b}$). The ``power'' $P(N_\text{s})$ is calculated
as:
\begin{equation}
 \int_{\lambda_{3\sigma}}^\infty g(\lambda | N_\text{s} + \text{b}) \text{d}\lambda = P(N_\text{s}).
\label{power}
\end{equation}

Let $n_{3\sigma}$ be the value of $N_\text{s}$ for which
$P(N_\text{s})=0.5$. Then $n_{3\sigma}$ is the number of expected signal
events that would lead to a detection with a significance of at least $3\sigma$
in 50\% of the cases.

If the analysis has been performed with a model for the source that predicts a
flux $\Phi_s$ and a mean number of signal events $\langle{n_s}\rangle$, the 
discovery potential will be given by
 \begin{equation}
\Phi_{3\sigma}= \Phi_{s}\cdot\dfrac{n_{3\sigma}}{\langle{n_{s}\rangle}}\,.
\label{eq:flux_calc}
\end{equation}
 
The sensitivity is calculated as the 90\% confidence level median upper limit by using the Neyman method \cite{Neyman}. 
The procedure is similar to that described previously but in this case a reference value $\lambda_{90}$ is calculated as the median of the $g(\lambda | \text{b})$ distribution. The power is evaluated as in Eq.~(\ref{power}) but with $\lambda_{90}$ instead of $\lambda_{3\sigma}$.
The number of events needed to reach the required sensitivity, $n_{90}$, is the number of events such that $P(N_\text{s})=0.9$.

As in Eq.~(\ref{eq:flux_calc}), the sensitivity flux $\Phi_{90}$ is calculated as:
\begin{equation}
\Phi_{90}= \Phi_{s}\cdot\dfrac{n_{90}}{\langle{n_{s}\rangle}}\,.
\label{eq:flux_calc_s}
\end{equation}

\subsection{Results}

\begin{figure}[h]
\includegraphics[width=.5\textwidth]{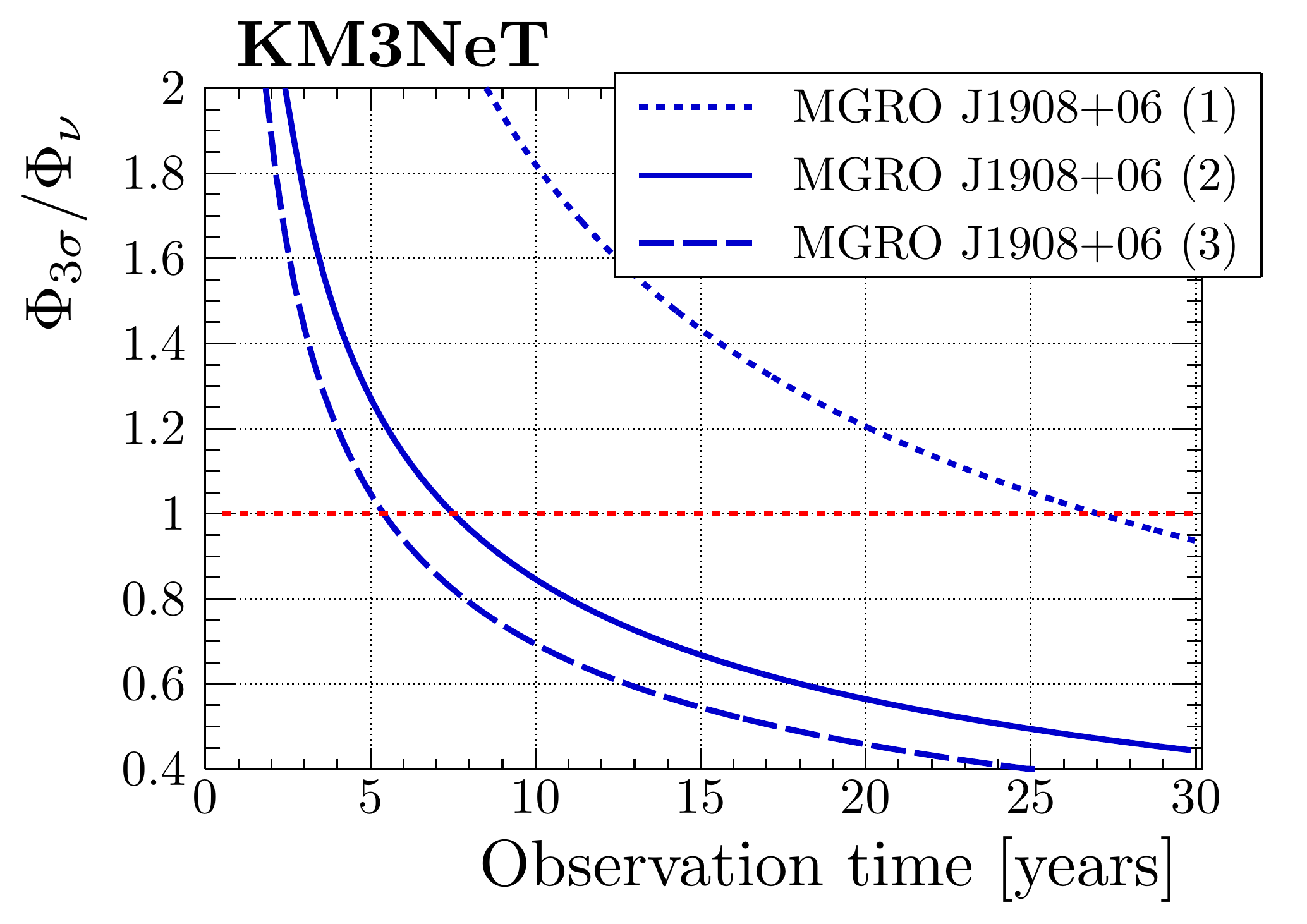}%
\includegraphics[width=.5\textwidth]{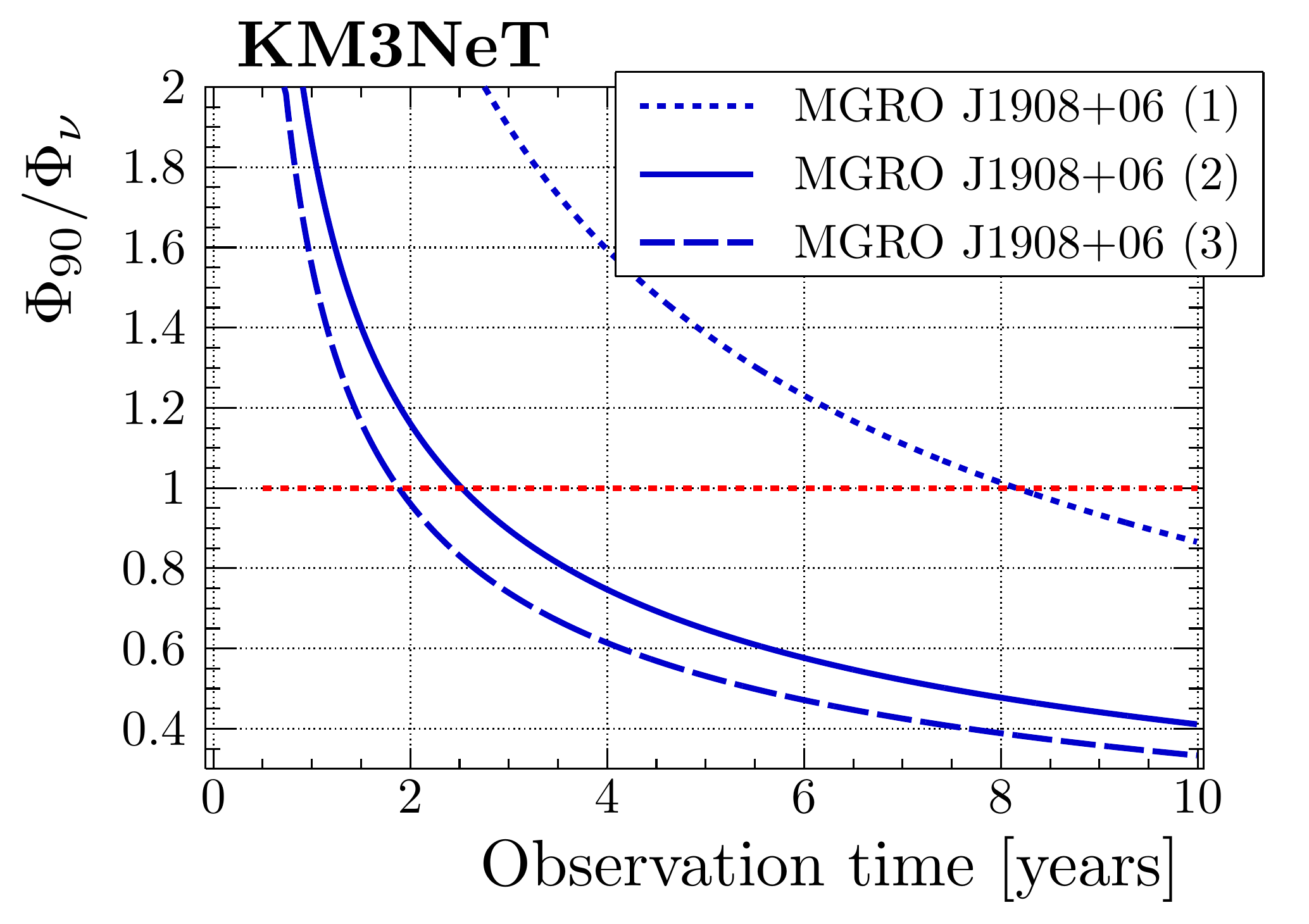}
\caption{
Ratio of the discovery potential $\Phi_{3\sigma}$ (left) and sensitivity
$\Phi_{90}$ (right) to the expectation flux $\Phi_{\nu}$ as a function of the observation time for the three
fluxes assumed for the source MGRO\,J1908+06 (see Table~\ref{tab:source}). The
fluxes (1), (2) and (3) correspond to a $\gamma$-ray spectral index
$\Gamma_\gamma = 2$ and cutoff energies of
$E_{\text{cut},\gamma}=30,300,800\,$TeV, respectively.}
\label{fig:mgro}
\end{figure}

\begin{figure}[h]
\includegraphics[width=.5\textwidth]{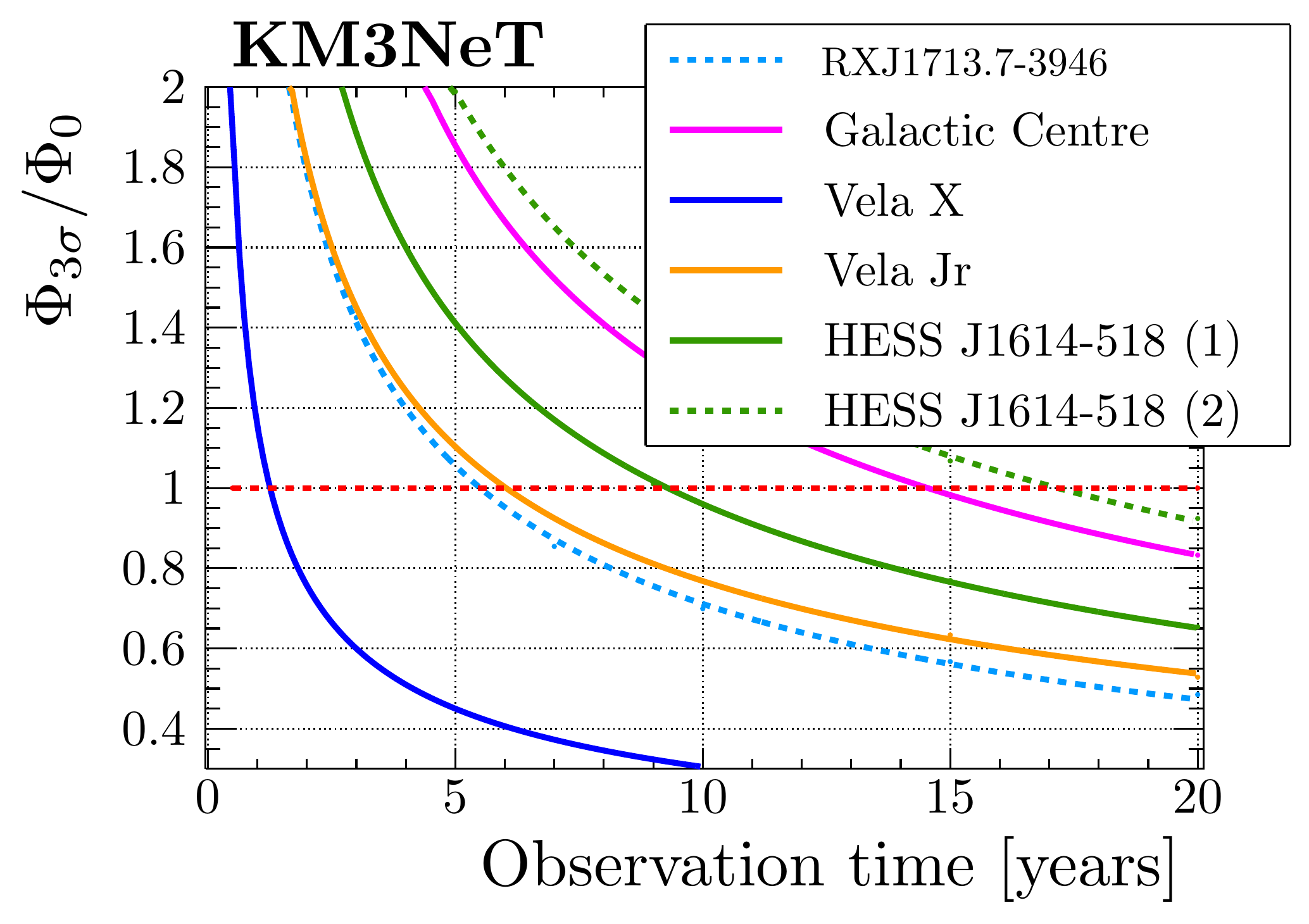}%
\includegraphics[width=.5\textwidth]{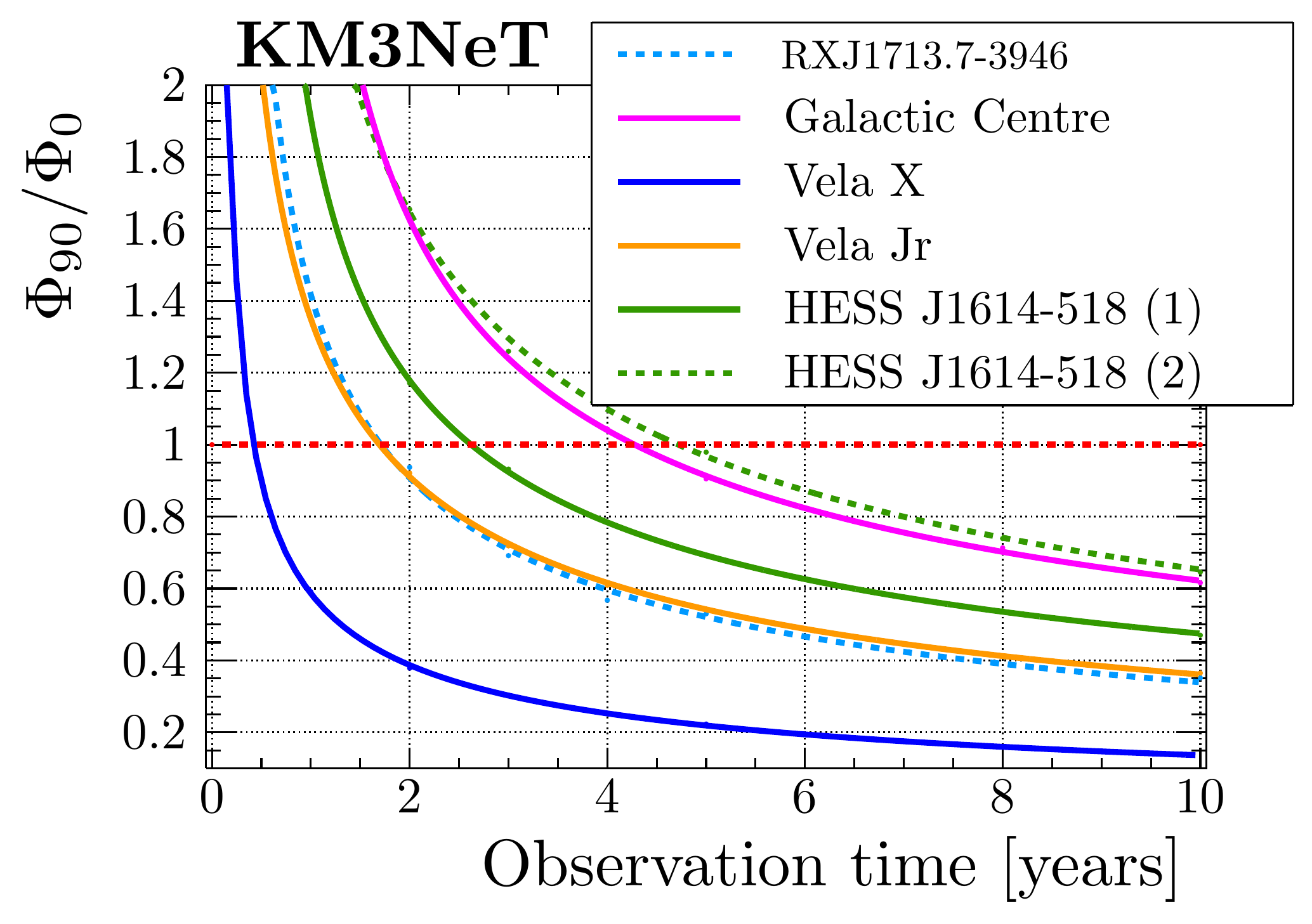}
\caption{
Ratio of the discovery potential $\Phi_{3\sigma}$ (left) and sensitivity
$\Phi_{90}$ (right) to the expectation flux $\Phi_{\nu}$ as a function of the observation time for the first
seven source fluxes listed in Table~\ref{tab:source}.}
\label{fig:galactic}
\end{figure}

The results are shown in Figs.~\ref{fig:mgro}, \ref{fig:galactic}. Figure
\ref{fig:mgro} refers to the source MGRO\, J1908+06 with the three neutrino flux
assumptions listed in Table~\ref{tab:source}, while the results for the other
sources are shown in Fig.~\ref{fig:galactic}. 
The fluxes corresponding to the discovery potential at 3$\sigma$, $\Phi_{3\sigma}$, are shown in the left plots of Figs.~\ref{fig:mgro} and \ref{fig:galactic} and the sensitivity at 90\% confidence level, $\Phi_{90}$, is shown the right plots.
In Figs.~\ref{fig:mgro}, \ref{fig:galactic} both $\Phi_{3\sigma}$ and $\Phi_{90}$ are reported as a ratio over the flux expectation $\Phi_{\nu}$ of each source,  given by Eq.~(\ref{eq:flux}) and shown in Table~\ref{tab:source}. 
Therefore $\Phi_{3\sigma}$/$\Phi_{\nu}$~=~1 indicates the time needed for a 3 sigma detection of the source for $\xi_\text{had}=1$. 
If $\Phi_{3\sigma}/\Phi_{\nu}<1$, $\Phi_{3\sigma}/\Phi_{\nu} = y$ gives the time needed to observe the source at $3\sigma$ for the case $\xi_\text{had}=y$. The same notation applies to the sensitivity.

Note that, by definition, $\Phi_{3\sigma}$ and $\Phi_{\nu}$ have the same spectral shape (see Eq.~(\ref{eq:flux_calc}) so the ratio between the two fluxes corresponds to the ratio between their normalisation constants and does not depend on the energy.

The sensitivity to exclude the predicted fluxes at 90\% confidence level is reached for all the
sources after about 5~(7) years for $\xi_\text{had}=1$ ($\xi_\text{had}=0.8$). 
For MGRO\,J1908+06, a $3\sigma$ discovery is possible after about 5.5\;years
(7.5\;years) if the cutoff in the $\gamma$-ray spectrum is
$E_{\text{cut},\gamma}=800$\,TeV~(300\,TeV) and if $\xi_\text{had}=1$. 
In the case of a cutoff at much lower energies, however, longer observation time would be necessary, e.g. 27 years for a cutoff at 30 TeV.
For RX\,J1713.7-3946, 5 years of
observation time are sufficient to constrain the hadronic fraction to
$\xi_\text{had}<0.5$. 
Even though hadronic scenarios for Vela~X are disfavoured,
it is worth noting that KM3NeT/ARCA could constrain the hadronic contribution to
$\xi_\text{had}<0.6$ ($\xi_\text{had}<0.2$) in about 1\,year (5.5\;years).

It should be noted that the results for a given neutrino flux degrade with increasing extension
of the source. This effect depends mainly on the source radius, but also on the
source spectrum. Studies concerning this effect have been reported by the ANTARES Collaboration
\cite{Antares9y, Antares2014} and other authors \cite{ambrogi}. To quantify the impact of the source extension,
the discovery potential and the sensitivity have been determined for two
sources, assuming that they are point-like instead of having finite extension. 
For RX\,J1713.7-3946, with a radius of $0.6^\circ$, $\Phi_{3\sigma}$ is
reduced by about 25\% and $\Phi_{90}$ by about 20\%. For MGRO\,J1908+06
($0.34^\circ$ radius and a harder spectrum), the relative reduction is about half as
large. 
Systematic effects from the
uncertainties of the source extensions or possible inhomogeneities of the
neutrino emission from the source region are expected to be negligible.

A stacking analysis has been performed for the two most intense SNRs, RX\,J1713.7-3946 and Vela Jr. 
The analysis is
similar to the one described above for single sources, with the PDFs in
Eq.~(\ref{eq_cap5_15}) obtained as weighted sums of the PDFs of the single
sources, both for the signal and the background, using as weight the number of
events expected in each case. 
Since these sources are quite distant in the sky (about $80^\circ$), there is no overlap between the selected events around the two sources. It is therefore possible to use the original classifier designed for each source as PDF for the stacked search.

\begin{figure}[h]
\includegraphics[width=.5\textwidth]{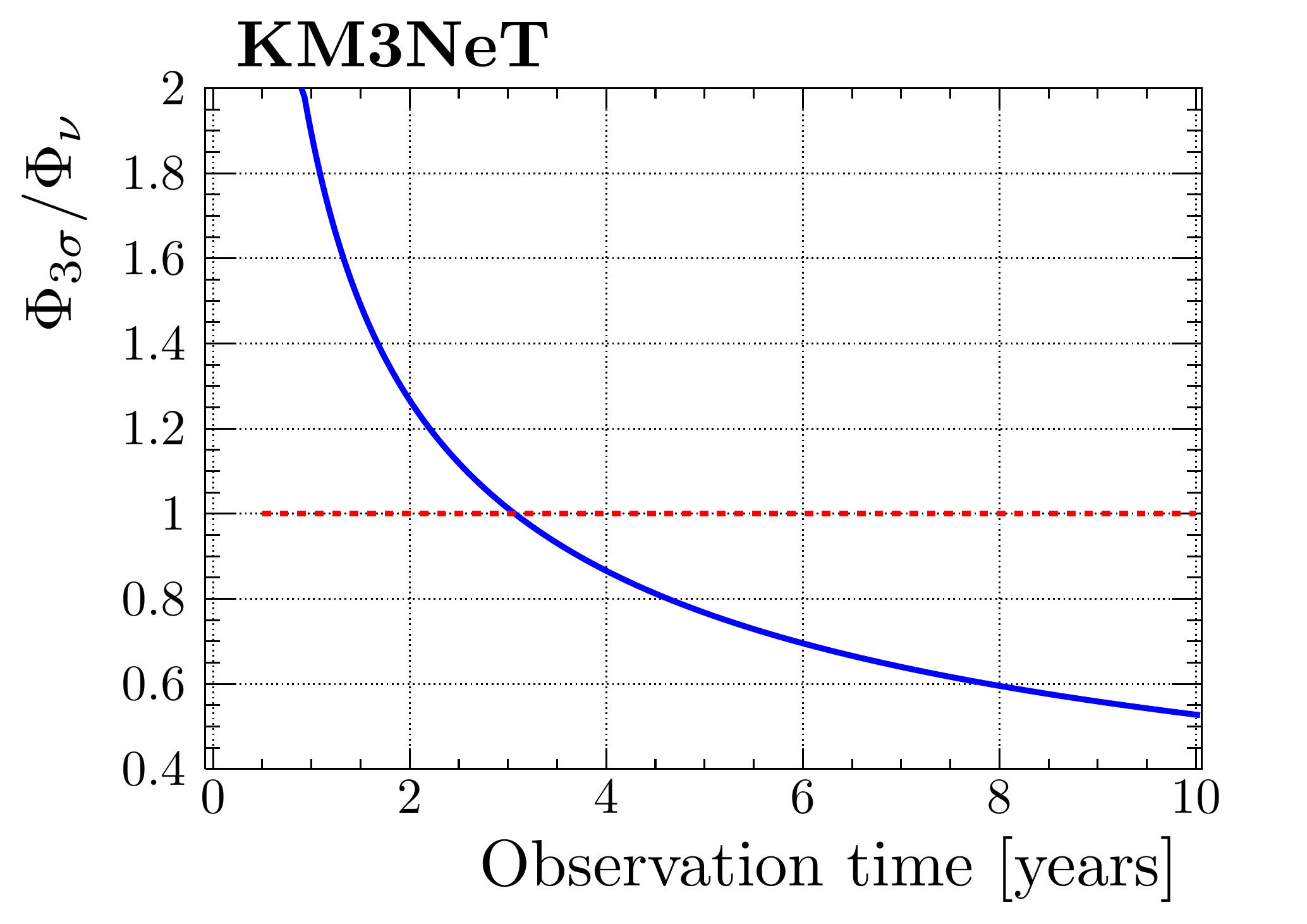}%
\includegraphics[width=.5\textwidth]{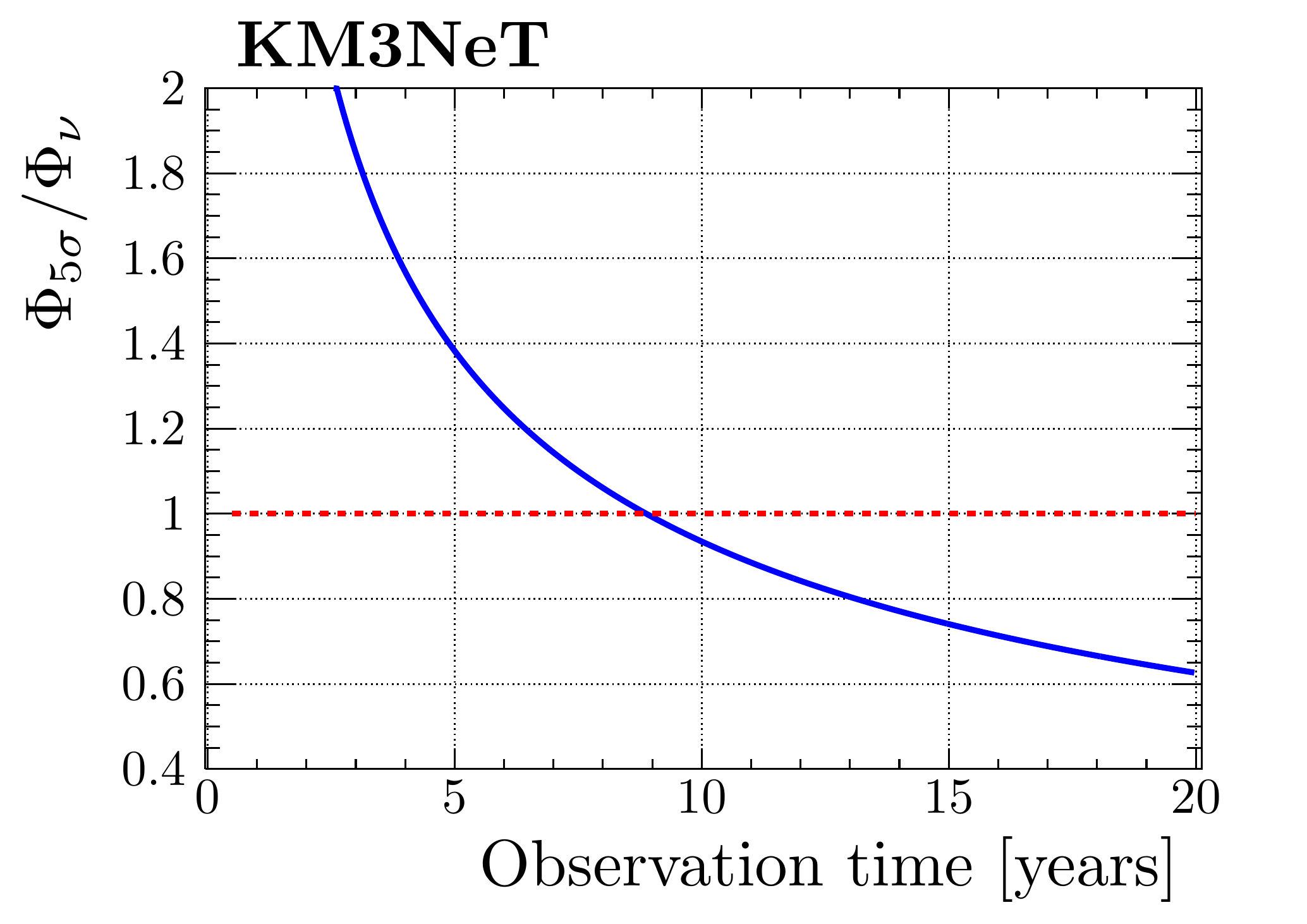}%
\caption{
Ratio of the discovery potentials $\Phi_{3\sigma}$ and $\Phi_{5\sigma}$ to the expectation flux $\Phi_{\nu}$ as a
function of the observation time for the stacking analysis including
RX\,J1713.7-3946 and Vela Jr. The neutrino fluxes assumed for the individual
sources are listed in Table~\ref{tab:source}. In this case, $\Phi_{\nu}$ is taken as the sum of the fluxes of the two sources.}
\label{fig:stack}
\end{figure}

In Fig.~\ref{fig:stack}, the resulting
values of $\Phi_{3\sigma}/\Phi_{\nu}$ and $\Phi_{5\sigma}/\Phi_{\nu}$ are shown as a
function of the observation time. Note that in this case $\Phi_{\nu}$ indicates the sum of the fluxes of the two stacked sources. An observation at 3$\sigma$ is possible after 3 years and at
5$\sigma$ after 9 years.

\section{Generic point sources with $E^{-2}$ spectrum}
\label{sec:E-2}

The sensitivity to astrophysical neutrino sources lacking a specific neutrino
flux prediction based on $\gamma$-ray measurements is performed assuming a
generic, unbroken power law energy spectrum proportional to $E^{-2}$. 
This assumption is in agreement with the recent IceCube findings \cite{IceCubeBlazar}
and provides a benchmark
scenario that can be compared with other detectors (see e.g.\ the
corresponding results from ANTARES \cite{Antares9y} and IceCube
\cite{IceCube7y}).

In this case no specific source generation is performed. Instead of a
specific training for each possible source location in the sky, only one
training is performed assuming as ``signal'' an event sample generated with a
$E^{-2}$ spectrum and imposing the experimental point spread function. Only
tracks reconstructed below the horizon and up to 10$^{\circ}$ above the horizon are considered. The
features used for the training are the same as for Galactic sources, except that in this case
the distance from the source position is not used at this stage of the analysis. The output of the Random Decision Forest classifier is used as a cut
variable in the analysis.

The likelihood ratio in Eq.~(\ref{eq_cap5_15}) is built in this case from the PDFs that describe the reconstructed directions and energies of the events, following a procedure widely used by the ANTARES Collaboration (see e.g. \cite{Antares9y}).
More precisely, 
\begin{equation}
\text{f}(x_{i}|\text{s})=  \text{f}(\psi_{i}|\text{s}) \, \text{f}(E_{i,\text{rec}}|\text{s})
\end{equation}
where $\text{f}(\psi_{i}|\text{s})$ is a
parameterisation of the point spread function, i.e.\ the probability density
function of reconstructing event $i$ at an angular distance $\psi_i$ from the
true source location, and $\text{f}(E_{i,\text{rec}}|\text{s})$ is the probability density function for signal events to be reconstructed with an energy $E_\text{rec}$. 
For the background, the
spatial part of the PDF depends only on the event declination $\delta_i$ while
the probability in right ascension is uniformly distributed, so
\begin{equation}
\text{f}(x_{i}|\text{b})=  \text{f}(\delta_i|\text{b})/(2\pi) \, \text{f}(E_{i,\text{rec}}|\text{b})
\end{equation}

Here, $\text{f}(\delta_i|\text{b})/(2\pi)$ is the probability density for background
events as a function of the declination and $\text{f}(E_{i,\text{rec}}|\text{b})$
is the probability density function for background events to be reconstructed with an energy $E_\text{rec}$. 

\begin{figure}[h]
\includegraphics[width=.5\textwidth]{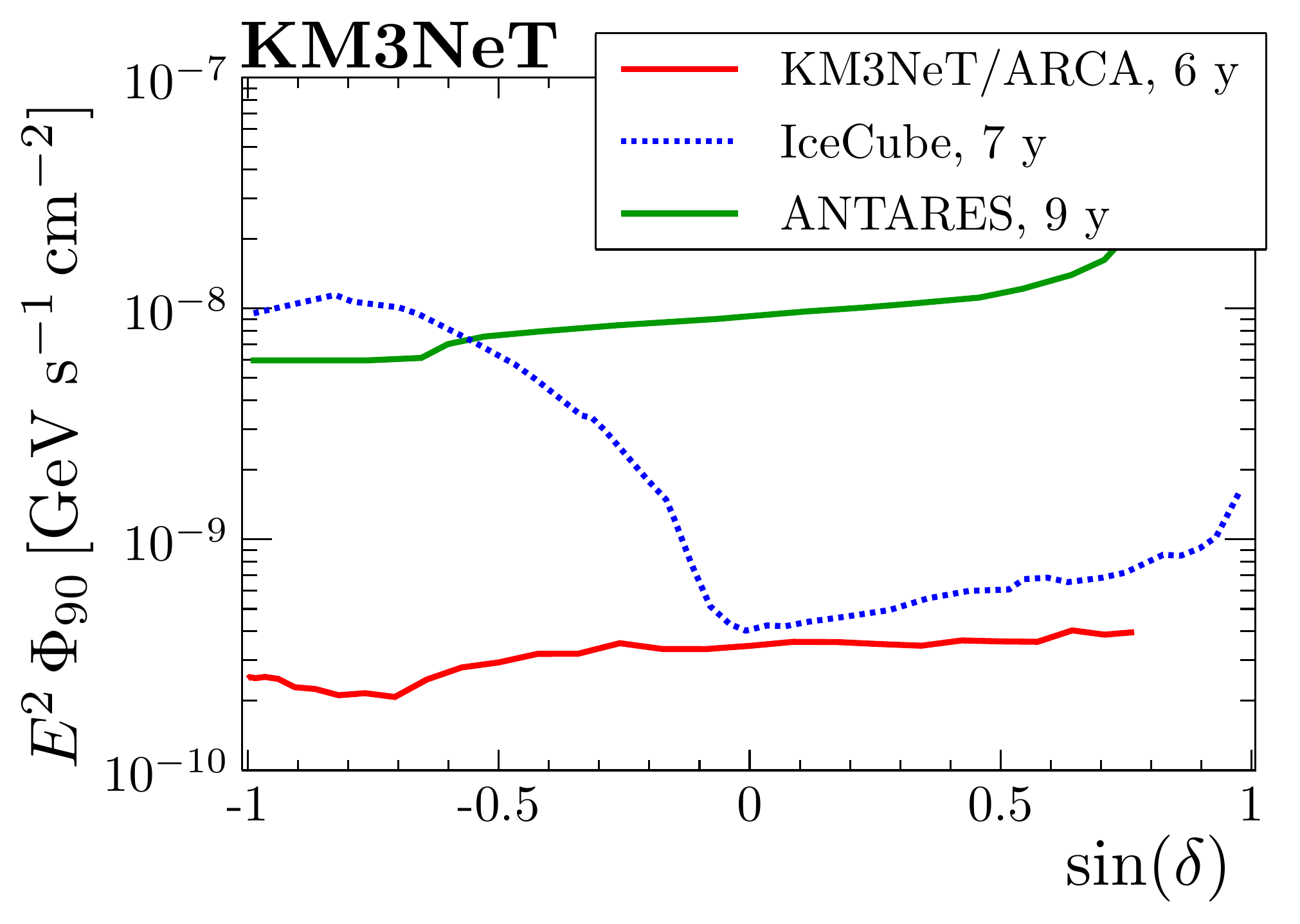}%
\includegraphics[width=.5\textwidth]{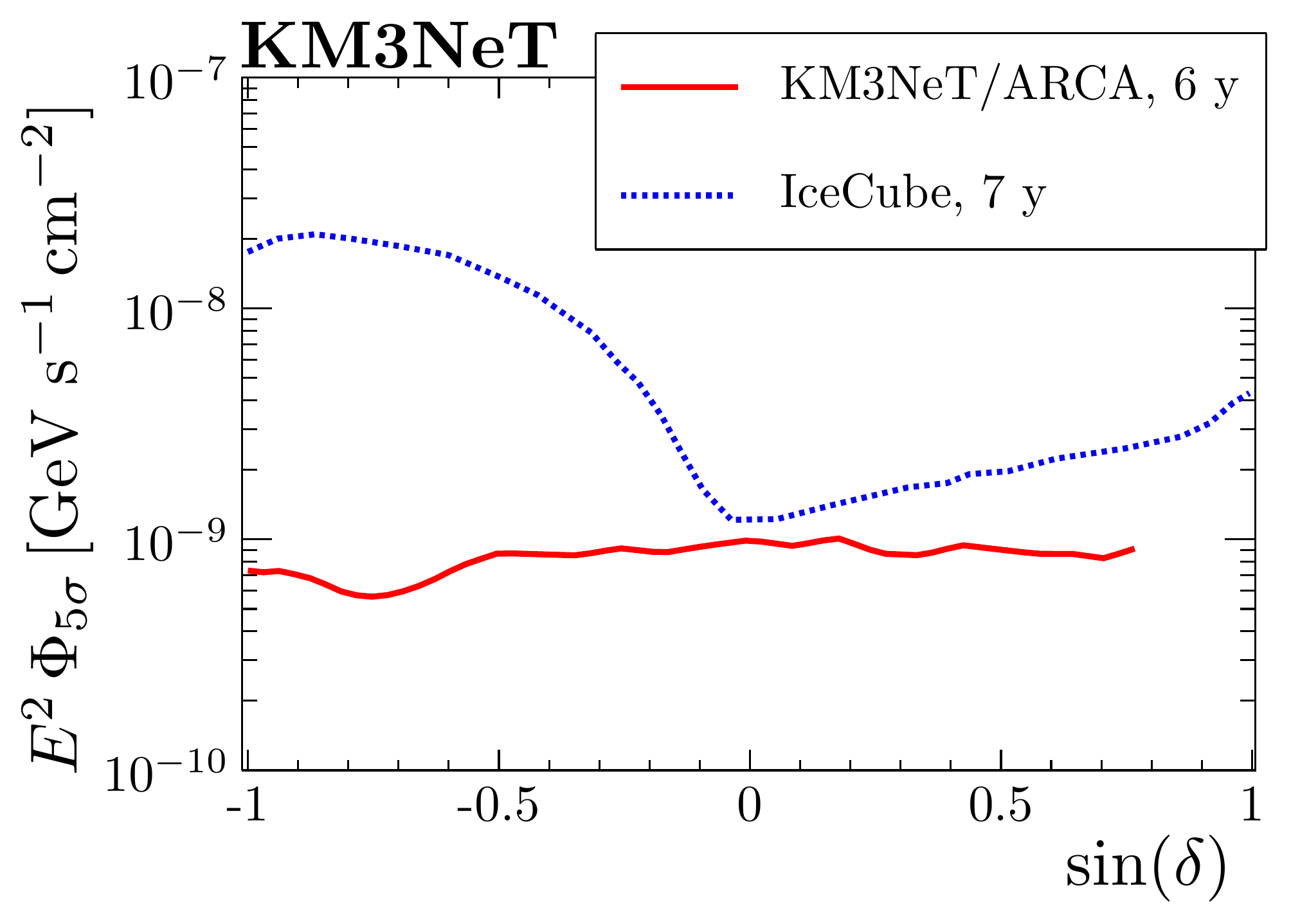} 
\caption{
Sensitivity, defined as the median upper limit at 90\% confidence level (left), and discovery flux at 5$\sigma$ (right) for sources with a
generic, unbroken neutrino flux proportional to $E^{-2}$, as a function of the
source declination. An observation time of 6\,years is assumed. For comparison,
the corresponding IceCube \cite{IceCube7y} and ANTARES \cite{Antares9y} results
are also shown.
Note that the IceCube discovery potential \cite{IceCube7y} follows the one-sided gaussian probability convention, while in this paper the two-sided one is used. For the KM3NeT results the difference deriving from using one or the other convention has been evaluated to be less than 4\%, within the line thickness of the figure.
}
\label{fig:E-2}
\end{figure}

The resulting sensitivity and 5$\sigma$ discovery flux are shown in
Fig.~\ref{fig:E-2} as a function of the source declination. 
An observation time of 6 years has been used, which is similar to IceCube results reported in Ref.\ \cite{IceCube7y}.

Previously \cite{loi}, the 5$\sigma$ discovery flux was reported for an
observation time of three years. 
The present analysis leads to a 25\% improvement with respect to Ref. \cite{loi} in the 5$\sigma$ discovery flux.

\section{Systematic uncertainties}
\label{sec:systematics}

A detailed investigation of the systematic effects for point source searches has
been reported in Ref.\ \cite{loi}. 
The main contribution comes from the
uncertainty on the normalisation of the conventional part of the atmospheric
neutrino flux, which is around $\pm25\%$ \cite{Honda}. The
corresponding variations of the discovery fluxes reported here are in the range
of about $+15\%$ to $-5\%$.

The uncertainties on the scattering and absorption lengths of Cherenkov light in the
deep-sea water and on the geometrical acceptance of the optical modules, which represents the major uncertainty in the response of a DOM to incident photons, have
negligible effects. Also, the deterioration of the event reconstruction due to
uncertainties on the position calibration is found to be small.

\section{Conclusions} 
\label{sec:conclusion}

The search for Galactic point-like neutrino sources is one of the prime goals of
the future KM3NeT/ARCA neutrino telescope. For a selected sample of Galactic sources
 the detection perspectives of KM3NeT/ARCA have been investigated, using several
parameterisations of the expected neutrino fluxes derived from the measured
$\gamma$-ray fluxes. 
A new event
reconstruction method \cite{Karel} and an improved multivariate analysis
allowing for the distinction of three event classes have been applied, improving upon the results of Ref.\ \cite{loi}.

Most of the Galactic sources considered can be observed by KM3NeT within a few
years if their $\gamma$-ray emission is of purely hadronic origin. As an
example, Vela Jr can be observed with a 3$\sigma$ significance within 6\,years,
and RX\,J1713.7-3946 within 5.5\,years. If no signal is observed after about 5 years,
the hadronic contribution to the $\gamma$-ray emission can be constrained to be less than 50\% for both sources.

The search for extragalactic neutrino sources is strongly motivated by the recent observation of a high-energy neutrino event coincident in direction and time with a $\gamma$-ray flaring state of a blazar \cite{IceCubeMultiMess}. 
In this respect, the performance of the KM3NeT/ARCA telescope have been investigated for a generic $E^{-2}$ neutrino flux.
The sensitivity and 5$\sigma$ discovery potential for sources with an unbroken
$E^{-2}$ spectrum for an observation time of 6\,years are in the ranges
$E^2\Phi=0.2\div0.4\times10^{-9}\,\text{GeV}\,\text{s}^{-1}\,\text{cm}^{-2}$
and $0.5\div1\times10^{-9}\,\text{GeV}\,\text{s}^{-1}\,\text{cm}^{-2}$,
respectively, for the full declination range $-1\le\sin(\delta)\lesssim0.8$. 
These values are similar to the results, based on a similar exposure, reported
by IceCube for the Northern hemisphere and by more than one order of magnitude
better for the Southern hemisphere.

\section{Acknowledgements}
The authors acknowledge the financial support of the funding agencies:
Agence Nationale de la Recherche (contract ANR-15-CE31-0020),
Centre National de la Recherche Scientifique (CNRS), 
Commission Europ\'eenne (FEDER fund and Marie Curie Program),
Institut Universitaire de France (IUF),
IdEx program and UnivEarthS Labex program at Sorbonne Paris Cit\'e (ANR-10-LABX-0023 and ANR-11-IDEX-0005-02),
France;
Deutsche Forschungsgemeinschaft (DFG),
Germany;
The General Secretariat of Research and Technology (GSRT),
Greece;
Istituto Nazionale di Fisica Nucleare (INFN),
Ministero dell'Istruzione, dell'Universit\`a e della Ricerca (MIUR),
Italy;
Ministry of Higher Education, Scientific Research and Professional Training,
Morocco;
Nederlandse organisatie voor Wetenschappelijk Onderzoek (NWO),
the Netherlands;
The National Science Centre, Poland (2015/18/E/ST2/00758);
National Authority for Scientific Research (ANCS),
Romania;
Plan Estatal de Investigaci\'on (refs.\ FPA2015-65150-C3-1-P, -2-P and -3-P, (MINECO/FEDER)), Severo Ochoa Centre of Excellence program (MINECO), Red Consolider MultiDark, (ref. FPA2017-90566-REDC, MINECO) and Prometeo and Grisol\'ia programs (Generalitat Valenciana), Spain.






\section{References}
\bibliographystyle{elsarticle-num} 
\bibliography{KM3NeT_PUB_2017_003-Point_like_ATrovato_v15}

\end{document}